%% file: t_report.tex
\documentclass[final,5p,times,twopage,twocolumn,a4paper]{elsarticle}
\usepackage{graphicx}
\usepackage{multirow}
\usepackage{amsmath,amssymb,amsfonts}
\usepackage{amsthm}
\usepackage{mathrsfs}
\usepackage[title]{appendix}
\usepackage{xcolor}
\usepackage{textcomp}
\usepackage{manyfoot}
\usepackage{booktabs}
\usepackage{algorithm}
\usepackage{listings}
\usepackage{threeparttable}

\usepackage[hyphens]{url}
\usepackage{babel}
\usepackage{adjustbox}
\usepackage{soul}
\usepackage{lscape}

\usepackage[frozencache,cachedir=.]{minted}

\usepackage{pbalance}   

\usepackage{worldflags}

\usepackage{hyperref}
\hypersetup{colorlinks,
  pdftitle={Tracing Data Packet Paths over the Internet using Traceroute},
  pdfauthor={Thomas Dreibholz, Somnath Mazumdar},
  pdfsubject={Internet, Traceroute},
  pdfkeywords={Internet Protocol, Packets, Path, Round-Trip Time, Traceroute, Ping},
  linkcolor=black,
  urlcolor=black,
  citecolor=black,
  filecolor=black
}
\usepackage{orcidlink}

\newcommand{\noun}[1]{\textsc{#1}}

\begin{document}

\begin{frontmatter}

\date{2023-02-01}
\journal{Technical Report}

\title{Tracing Data Packet Paths\\over the Internet using Traceroute}

\author{Thomas Dreibholz\orcidlink{0000-0002-8759-5603}}
\address{Centre for Resilient Networks and Applications, Simula Metropolitan \\
Stensberggata 27, 0170 Oslo, Norway \\
\href{mailto:mailto:Thomas Dreibholz (托马斯博士) <dreibh@simula.no>}{dreibh@simula.no} \\
\url{https://www.simula.no/people/dreibh}}

\author{Somnath Mazumdar\orcidlink{0000-0002-1751-2569}}
\address{Department of Digitalization, Copenhagen Business School, \\ Solbjerg Plads 3, 2000 Frederiksberg, Denmark \\
\href{mailto:mailto:Somnath Mazumdar <sma.digi@cbs.dk>}{sma.digi@cbs.dk} \\
\url{https://research.cbs.dk/en/persons/somnath-mazumdar}}

\begin{abstract}
Network communication using the Internet Protocol~(IP) is a pillar of modern Internet applications. IP allows data packets to travel the world through a complex set of interconnected computer networks managed by different operators. How IP-based data communication changes over time can be interesting from an end-system's perspective without relying on underlying network providers. This article presents an extensive, trace-driven analysis of user data traffic (covering five years of observations, six large Internet service providers (covering research, business and consumer category type), twenty autonomous systems, and fourteen countries.

Our three primary findings are:
\textit{i)} users data packet transmission paths are not deterministic and does not always select the geographically shortest path;
\textit{ii)} user packets take different routes that cover many countries and detour between two fixed points. Even after changing the types of Internet service provider type (e.g., from commercial to research), the routing can differ significantly between two locations.
\textit{iii)} Packet transmission delay can be influenced by changing the Internet service provider and IP protocol versions (i.e., from IPv4 to IPv6).
\end{abstract}


\end{frontmatter}

\section{Introduction}
Today's Internet infrastructure is complex and very dynamic. A data packet routing path is typically a combination of Internet service provider (ISP) and autonomous system~(AS) paths. Traditionally, ISPs use a combination of local policies that include commercial relationships~\cite{Mahajan2005negotiation,FTC2021}, the length of AS paths~\cite{Spring2003causes}, and resource constraints to select data packets path(s). ISPs use Border Gateway Protocol~(BGP) as the inter-domain routing protocol to exchange path information~\cite{RFC4271}. It has been found that inter-domain routing~\cite{Mahajan2005negotiation,Gao2016quantifying,NoF2014} and peering policy can impact path length~\cite{Spring2003causes,Gao2016quantifying,Wang2022traffic}. 
In general, packet forwarding policies affect the performance of web applications and services.

\subsection{Research Context}
Given this scenario, our primary research question here is \textit{How user data packets travel on the Internet today?} We formulate this research problem from an end-user perspective where no information or cooperation from the underlying network providers is expected. To analyze the question in detail, we divided it into two research sub-questions: \textit{i) How ISP allocate the data traffic route between two fixed points?} and \textit{ii) How does network performance get affected while data are in transit?} We answered these two research sub-questions using a trace-based analysis of data collected over five years (from~2018 to~2022)\footnote{We lost the measurement sites in China and Germany by the end of~2022, due to tightened firewall policies.}. The analysis spans over 14~countries, using six large ISPs, including 20~different autonomous systems~(AS).

\subsection{Research Findings}
We list our three main findings to answer two research sub-questions:
\begin{enumerate}
    \item \textbf{[For first sub-question]} Even between two fixed points, user packets take different routes covering many countries and detours (refer to Subsubsection~\ref{subsub:Traceroute-KAU-NTNU}). It does not always consider the shortest geographic distances. We found that the ISPs of a research network may include more third-party countries than the commercial ISPs (see Subsubsection~\ref{subsub:Traceroute-HU-NTNU}).
    \item \textbf{[For first sub-question]}  We also found that ISPs of the same category type prefer ISPs of the same category type (i.e.\ research network or commercial network; see Subsection~\ref{subsub:Traceroute-UDE-NTNU}).
    \item \textbf{[For second sub-question]}  Network latency can be influenced by changing the ISPs and the IP protocols from IPv4 to IPv6 (refer to Subsubsection~\ref{subsub:Traceroute-UDE-NTNU}).
\end{enumerate}
To support reproducibility, we publicly released the code as open-source software under GPL~3.0 license.

\section{Related Work}
Traceroute is a popular tool to discover the Internet topology~\cite{Donnet2007internet} but has some limitations. Overall, it is one of the well-supported, less intrusive approaches to identifying routing issues that can impact users and applications~\cite{Maier2023loop,Luttringer2019let,Marchetta2013drago,Donnet2007internet,Paxson1997end}. Below, the papers are mentioned following chronological reverse-order.
\subsection{Path Inflation and Routing Policies}
Multiple pieces of literature focus on path inflation~\cite{Wang2022traffic,Gao2016quantifying,Spring2003causes} and routing policies~\cite{Gao2016quantifying,NoF2014,Mahajan2005negotiation}. Wang et al.\ focused on the path inflation ratio using the IP address geo-location method to detect the points of presence where clouds exchange traffic with the rest of the Internet~\cite{Wang2022traffic}. Gao et al.\ measured the AS path lengths and found that AS paths are inflated due to inter-domain routing policies, where the shortest AS path routing policy is not used in most cases~\cite{Gao2016quantifying}. Spring et al.~used traceroute data collected from PlanetLab and found potential sources of path inflation~\cite{Spring2003causes}.

Golkar et al.~compared the routing of IPv4 and IPv6, showing that IPv6 routes can change more frequently~\cite{NoF2014}. At the same time, the number of hops is similar to the IPv4 routing. Mahajan et al.~presented a negotiation framework to share information among ISPs about data traffic flows based on a specific relationship~\cite{Mahajan2005negotiation}.

\subsection{Path Tracing}
Alaraj et al.\ studied routing loops (between two vantage points) to all IPv4 addresses by traceroutes to a limited range of Time-to-Lives~(TTLs)~\cite{Alaraj2023global}, while Maier et al.\ studied persistent routing loops considering IPv4 and IPv6 and found routing loops are still a matter of concern~\cite{Maier2023loop}. Vermeulen et al.\ proposes a mechanism to run reverse path measurements from source to destinations, which was meant for operators for better traffic engineering~\cite{Vermeulen2022internet}, while Giotsas et al.\ proposes a method to use traceroutes for long periods of time and to avoid wasting measurements on unchanged paths using the RIPE Atlas traceroute corpus that becomes stale due to path changes~\cite{Giotsas2020reduce}. The aim of this study was to identify all border IP changes. In another study, Luttringer et al.\ detects all multi-protocol label switching tunnels along a route using traceroute and ping probes~\cite{Luttringer2019let}, while Morandi et al. developed a tool to identify application flows to probe and of tracing the paths of multiple concurrent flows of both TCP and UDP flows~\cite{Morandi2019}. Marchetta at al.\ proposed a technique to quantify and locate hidden routers in traceroute IP paths~\cite{Marchetta2013drago}.

Our analysis found that the data packet transmission path is not deterministic, and data paths do not always select the geographically shortest path (similar to~\cite{Gao2016quantifying}); thus, significant detours are possible. In addition, packet transmission delays can be influenced by changing the network providers or IP protocol versions (similar to~\cite{NoF2014}).

\section{Methodology}
We are interested in knowing how data packets flow from source to destination without knowing ISP's infrastructure details. Our primary goal was to obtain knowledge about network communication and packet flow from an end user perspective. We did not measure between routers in the Internet because this would mean running software somewhere in the ISPs' networks. Our endpoints are universities that offer consumer-grade secondary ISP subscriptions.

The BGP analysis is clearly beyond the scope of this paper because the measurements are run from an end-user perspective. We do not rely on ISPs (or need cooperation from ISPs) to provide BGP data and measure traffic between the end systems. In this section, we provide a brief overview of the methodologies used, including the applied tools, measurement infrastructure, and dataset descriptions.

\subsection{Tools: Ping, Traceroute and \noun{HiPerConTracer}}
\subsubsection{Ping}
Ping (Packet Inter-Network Groper) refers to the well-known Unix program \texttt{ping}, which uses simple test messaging of the Internet Control Message Protocol~(ICMP)~\cite{RFC792} for IPv4 and ICMPv6~\cite{RFC4443} for IPv6 to test connectivity. The tool sends an \texttt{ICMP/ICMPv6 Echo Request} packet to a remote system, which then responds with an \texttt{ICMP/ICMPv6 Echo Reply} packet. The time between sending the \texttt{Echo Request} and \texttt{Echo Reply} denotes the Round-Trip Time~(RTT). RTT is the time it takes for a message to reach the destination and receive a response.

\subsubsection{Traceroute}
Traceroute is provided by the Unix \texttt{traceroute} command. It uses the same principle as Ping, but sends multiple packets (e.g.\ \texttt{ICMP/ICMPv6 Echo Requests}\footnote{Using UDP packets is also possible, and performed by various implementations. For the intended tracing purpose, the protocol of the sent packets makes no difference. However, since UDP is blocked more frequently in end networks, we had chosen ICMP.}). In each packet sent in a run, the IPv4 Time-to-Live~(TTL) or IPv6 Hop Limit counter is incremented. This counter is decremented by each router that forwards the packet. When it reaches zero, the packet is dropped without arriving at the actual destination, and an \texttt{ICMP/IMCPv6 Time Exceeded} error is generated. The source of this error message is the IP address of the router. Once arrived at the actual destination, an \texttt{Echo Request} is answered by an \texttt{ICMP/IMCPv6 Echo Response}. By collecting the \texttt{Time Exceeded} messages, and the final \texttt{Echo Reply}, information about all routers to the destination can be obtained.

For security reasons, the rate of sent \texttt{Echo Replies} and \texttt{Time Exceeded} is limited. Sometimes, they are even completely blocked by firewalls. Therefore, there is the possibility that not all routers respond. \texttt{ping} and \texttt{traceroute} are basic shell commands that are used to troubleshoot network connectivity interactively. Consequently, the extended execution of Ping and Traceroute measurements, with high frequency over IPv4 and IPv6, with support for multiple addresses (i.e.\ different ISPs) and parallelization, is provided by the \noun{HiPerConTracer} tool\footnote{High-Performance Connectivity Tracer~(HiPerConTracer): \url{https://www.nntb.no/~dreibh/hipercontracer/}.}~\cite{SoftCOM2020-HiPerConTracer}.
Each \emph{source} address is handled independently, i.e.\ measurements from different sources addresses can run in parallel as own threads. Each source address belongs to a different ISP subscription; IPv4 and IPv6 are independent protocols. All subscriptions have fixed public IP addresses. For Traceroute, each \emph{destination} address is probed sequentially, to avoid triggering ICMP rate limitations in routers (particularly near the source). Since we have all measurement endpoints under our own control, Ping measurements to all destinations can run as a burst, since our endpoints are not rate-limited.

\subsubsection{\noun{HiPerConTracer}}
\noun{HiPerConTracer} also handles the load-balancing issue by specially crafting packets of a run to take the same route. Load balancing typically uses the first 4~bytes of the Transport Layer header to map a packet onto the possible outgoing paths. For UDP and TCP, these bytes contain the source and destination port numbers. This prevents packets of the same connection from going over different paths, which may lead to reordering and poor performance due to the assumed packet loss by a protocol's congestion control. For the ICMP case, these 4~bytes contain the 2-byte ICMP checksum~\cite{RFC1071} and a fixed type (here: Echo Request or Echo Reply) and code field (here: 0). Therefore, to keep the 4~bytes constant, to avoid load balancing, the payload of the ICMP packets must be crafted to keep the checksum constant. In addition, \noun{HiPerConTracer} sends all packets of the TTL/Hop Limit sequence as a burst, minimizing the chance of route changes affecting a single Traceroute measurement run.

\subsection{Measurement Infrastructure}
\label{sec:measure_infra}
Traceroute provides an option to obtain used routes without any support or information provided by the ISPs involved in the routing. In general, routers support the underlying protocols used for Traceroute (i.e.\ ICMP and ICMPv6). Even if some ISPs may reveal routing history data, obtaining data from all (or at least most of) the ISPs  is doubtful. Therefore, due to the large number of ISPs, organizations, and countries involved in routing, Traceroute is one of the best ways to collect routing information based on the actual perspective of the end user. In addition to Traceroute observations, we performed fine-granular RTT measurements using ping.

The challenge of obtaining long-term routing information is to collect these data with the following properties:
\begin{enumerate}
 \item The actual routing information about hops (by \noun{HiPerConTracer} Traceroute);
 \item The RTT of end-to-end-communication data (by \noun{HiPerConTracer} Ping);
 \item With a high frequency of approximately every 300~s for routing and 1~s for RTT (to have a fine monitoring granularity to see changes over short time periods);
 \item Over a long period of 5~years (to also see changes over long time periods);
 \item Taking the application of load balancing into account (and prevent distorting the results by packets of a single probing measurement taking different routes);
 \item Using IPv4 and IPv6 wherever available (in order to allow for comparing both protocols),
 \item In different network types, i.e.\ research ISPs, business-grade ISPs, consumer ISPs (to see differences for varying network types).
\end{enumerate}


\subsection{Data Set Description}
We use \noun{HiPerConTracer} to obtain data focusing on \textit{how is a user data packet routed?} and \textit{how large is the RTT?} with the requested frequency and handling the load balancing (by adjusting the content, to prevent different routes of packets belonging together). The measurement data are imported by the \noun{HiPerConTracer} importer into a MongoDB database, using JavaScript Object Notation~(JSON) format.

\begin{listing}
\begin{minted}{json}

{
   "checksum": 64072,
   "destination": "195.159.158.134",
   "rtt": 48714,
   "source": "132.252.152.105",
   "status": 255,
   "timestamp": 1546300833842802
}
\end{minted}
\caption{Sample Ping example.}
\label{lst:Ping-Example}
\end{listing}
Listing~\ref{lst:Ping-Example} provides a \noun{HiPerConTracer} Ping example, with the content: absolute timestamp (microseconds since the Epoch\footnote{Unix epoch: January 1, 1970, 00:00:00.000000 UTC}), source and destination address (decoded as strings; binary storage), status (e.g., 255 = Echo Reply received), RTT (in microseconds).

\begin{listing}
\begin{minted}{json}

{
   "checksum": 63328,
   "destination": "2001:700:300:2211::128",
   "hops": [
      {
         "hop": "2001:638:501:4ef2::1",
         "rtt": 1900,
         "status": 1
      },
      {
         "hop": "2001:638:501:4fef::1",
         "rtt": 1243,
         "status": 1
      },
      {
         "hop": "2001:638:501:4f02::2",
         "rtt": 1286,
         "status": 1
      },
      ...
      {
         "hop": "2001:700:300::2e09",
         "rtt": 39630,
         "status": 1
      },
      {
         "hop": "2001:700:300:2211::128",
         "rtt": 39715,
         "status": 255
      }
   ],
   "pathHash": 4408508201904120687,
   "round": 0,
   "source": "2001:638:501:4ef2::105",
   "statusFlags": 768,
   "timestamp": 1546300978554768,
   "totalHops": 17
}
\end{minted}
\caption{One Traceroute example.}
\label{lst:Traceroute-Example}
\end{listing}
Similarly, an example for a \noun{HiPerConTracer} Traceroute run is provided in Listing~\ref{lst:Traceroute-Example}. In addition to Ping, it also contains the information about each hop, including its address (in binary form; here shown decoded as string), RTT, and status (e.g.\ 1 = Time Exceeded received; 255 = Echo Reply received).

\section{Trace Analysis}
\label{sec:Results}

\subsection{Measurement Setup}
\label{sub:Testbed-Setup}

\begin{table}
\caption{The \noun{NorNet Core} sites used.}   
\label{tbl:NorNet-Core-Sites}
\begin{center}
\begin{adjustbox}{max width=\columnwidth}
\small
\begin{threeparttable}
\begin{tabular}{|l|c|c|}
\hline
Site (Country) & ISP 1 & ISP 2 \\
\hline\hline
NTNU Trondheim (\worldflag[width=6pt,length=10pt,stretch=1]{NO}NO)             & \href{https://www.uninett.no}{Uninett}\tnote{F}   & \href{http://www.powertech.no}{PowerTech}\tnote{A}  \\ \hline
Karlstads Universitet (\worldflag[width=6pt,length=10pt,stretch=1]{SE}SE)      & \href{https://www.sunet.se}{SUNET}\tnote{2,F}     & – \\ \hline
Universität Duisburg-Essen (\worldflag[width=6pt,length=10pt,stretch=1]{DE}DE) & \href{https://www.dfn.de}{DFN}\tnote{F}           & – \\ \hline
Hainan University (\worldflag[width=6pt,length=10pt,stretch=1]{CN}CN)          & \href{http://www.cernet.edu.cn}{CERNET}\tnote{F}~ & \href{http://www.chinaunicom.com}{China Unicom}\tnote{1,C}~~~ \\ \hline
\end{tabular}
\begin{tablenotes}
\item[1] Only IPv4; IPv6 is not available from ISP.
\item[F] Research network fibre.
\item[C] Consumer-grade fibre.
\item[A] Consumer-grade Asymmetric Digital Subscriber\\Line~(ADSL).
\end{tablenotes}
\end{threeparttable}
\end{adjustbox}
\end{center}
\end{table}
For this purpose, we use the NorNet infrastructure\footnote{NorNet: \url{https://www.nntb.no}.}~\cite{ComNets2013-Core} to perform the \noun{HiPerConTracer} Ping and Traceroute measurements at different locations in multiple countries with various ISPs. Specifically, we collected Ping and Traceroute results from the year~2018 to the year~2022, between the sites shown in Table~\ref{tbl:NorNet-Core-Sites}. This infrastructure consists of distributed sites in Norway~(\worldflag[width=6pt,length=10pt,stretch=1]{NO}NO), Sweden~(\worldflag[width=6pt,length=10pt,stretch=1]{SE}SE), Germany~(\worldflag[width=6pt,length=10pt,stretch=1]{DE}DE), and China~(\worldflag[width=6pt,length=10pt,stretch=1]{CN}CN).

\noun{HiPerConTracer} Ping has been performed every second for each relation (available IP version, local ISP, remote ISP), i.e.\ around 157.5~million measurements per relation, measuring the time between sending an \texttt{ICMP Echo Request} and receiving the corresponding \texttt{ICMP Echo Response}. \noun{HiPerConTracer} Traceroute has been performed every approximately five minutes with three runs per relation (i.e.\ around 1.6~million measurements per relation).
Our measurements only create a relatively small load. The Ping and Traceroute packets are small (44~B for IPv4, 64~B for IPv6; including IP headers). Since our focus is on the traffic paths from a user's perspective, workload experiments (such as analyzing congestion control behavior) are out of scope.

IP addresses do not provide any geo-location-related information for the network. Accurate geo-location is a challenge~\cite{ANRW2020}, and to approximate the geo-location of a router, we use \noun{HLOC}~\cite{HLOC2017} for location approximation by performing \noun{RIPE Atlas} Ping measurements to obtain the RTT to the router's address from known vantage points. We took the HLOC results when their estimated distance was $\le$25~km. Furthermore, we use online queries to the \noun{ipinfo.io}\footnote{\noun{ipinfo.io}:~\url{https://ipinfo.io}.} service when HLOC is unable to provide a suitably accurate location.

Finally, we use AS information from the~\noun{CIDR Report}\footnote{\noun{CIDR Report} AS list: \url{https://www.cidr-report.org/as2.0/autnums.html}.} and look up AS numbers in the free \noun{GeoLite2}\footnote{\noun{GeoLite2}: \url{https://dev.maxmind.com/geoip/geoip2/geolite2/}.} database. That is, the mapping of IP~addresses to ASs is based on the data provided by the Regional Internet Registries~(RIR). We selected three scenarios for further analysis.
These scenarios are as follows:
\begin{enumerate}
 \item Between neighbor countries: Karlstad, \worldflag[width=6pt,length=10pt,stretch=1]{SE}SE to Trondheim, \worldflag[width=6pt,length=10pt,stretch=1]{NO}NO;
 \item Intra-continental: Essen, \worldflag[width=6pt,length=10pt,stretch=1]{DE}DE to Trondheim, \worldflag[width=6pt,length=10pt,stretch=1]{NO}NO;
 \item Inter-continental: Haikou, \worldflag[width=6pt,length=10pt,stretch=1]{CN}CN to Trondheim, \worldflag[width=6pt,length=10pt,stretch=1]{NO}NO.
\end{enumerate}

\subsection{Connectivity and Routing Findings}
\label{sub:Connectivity-Findings}
First, we look at the end-user perspective: RTTs from \noun{HiPerConTracer} Ping. From the results, we can conclude:
\begin{enumerate}
 \item The data traffic route changes over time;
 \item RTTs may change significantly, depending on the choice of ISPs. It is true even between neighboring countries;
 \item IPv4 and IPv6 results may differ significantly.
\end{enumerate}

\subsubsection{Scenario: Between Neighboring Countries}
\label{subsub:Ping-KAU-NTNU}

\begin{figure*}
\begin{center}
\includegraphics[width=1.00\textwidth]{%
   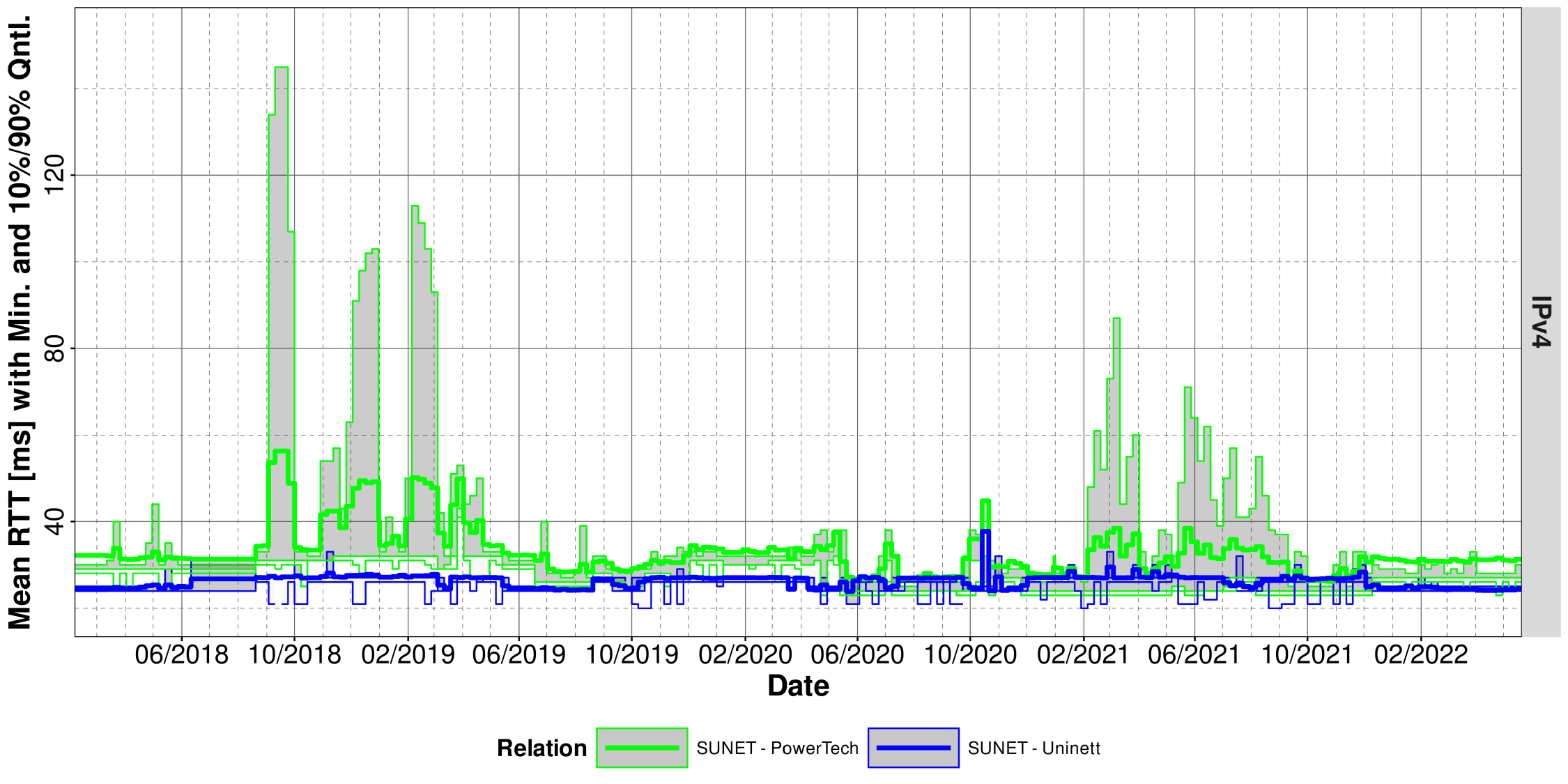}
\end{center}
\caption{RTT Time Series for Karlstad, SE to Trondheim, NO. Blue colour: high-speed fibre, green colour: ADSL connection.}
\label{fig:TimeSeries-KAU-NTNU}
\end{figure*}
In the beginning, we examine the RTT between two neighbouring countries: Karlstad, \worldflag[width=6pt,length=10pt,stretch=1]{SE}SE, and Trondheim, \worldflag[width=6pt,length=10pt,stretch=1]{NO}NO. The Karlstad site only has IPv4 connectivity through its only ISP SUNET (Swedish research network). In contrast, the site in Trondheim is connected to Uninett (Norwegian research network) and PowerTech (consumer ADSL subscription). Figure~\ref{fig:TimeSeries-KAU-NTNU} presents the RTT time series with the average RTT.
Here, the x-axis presents the long-term behavior of the RTT.
Furthermore, the difference between 10\%~and 90\%~quantiles is shown as a darker area, while the difference between the minimum RTT and 10\%~quantile is shown as a lighter area. The average and quantiles are computed over buckets of 1~hour. These buckets show the short-term behavior of the RTT.
We present the quantiles, since they are more interesting for service quality than just a median value: the bottom part (10\%), related to the best achievable RTTs, and the upper part (90\%), to show the typically achievable RTT range.

The RTT is relatively stable for transmission to Uninett, with only a slight variation during the observation time of almost five years. The average RTT is higher for Power\-Tech, which is an ADSL consumer network, and therefore expected. However, there are also significant peaks, particularly in~2018 with over 100~ms for the 90\%~quantile, and in~2021 with still over 60~ms.

\begin{figure*}
\begin{center}
\includegraphics[width=1.00\textwidth]{%
   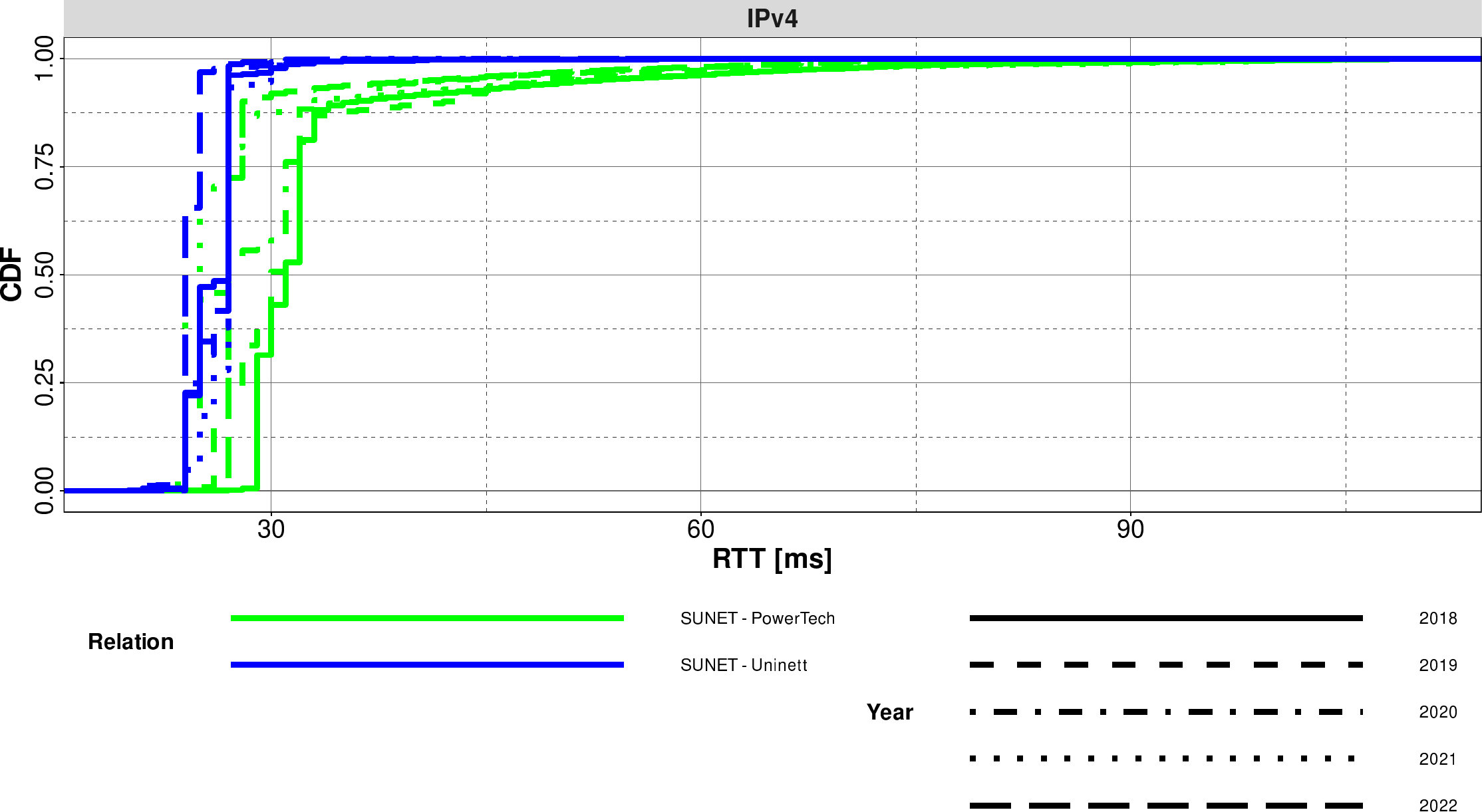}
\end{center}
\caption{RTT CDF for Karlstad, SE to Trondheim, NO.}
\label{fig:CDF-KAU-NTNU}
\end{figure*}
To further visualise the RTT differences, Figure~\ref{fig:CDF-KAU-NTNU} provides a cumulative distribution function~(CDF) plot for the mean RTT, showing the fraction of RTT measurements (y-axis) up to the given RTT value (x-axis). Different years use different line styles. We see more steps related to more significant changes in a year. As expected, the results for Uninett (in blue) are relatively stable, while PowerTech (in green) shows significantly more changes in each year. 2018~shows the highest average RTTs, as expected (solid, in green). From these results, we can conclude that the data packet route changes over the time. This occurs even between neighboring countries using different network mediums. That is, copper line (ADSL) and fibre cable may have changing performance over time. Next, we want to analyze: \textit{Is the data packet route for non-neighboring countries also experiencing similar performance changes over time?}

\subsubsection{Scenario: Intra-Continental}
\label{subsub:Ping-UDE-NTNU}

\begin{figure*}
\begin{center}
\includegraphics[width=1.00\textwidth]{%
   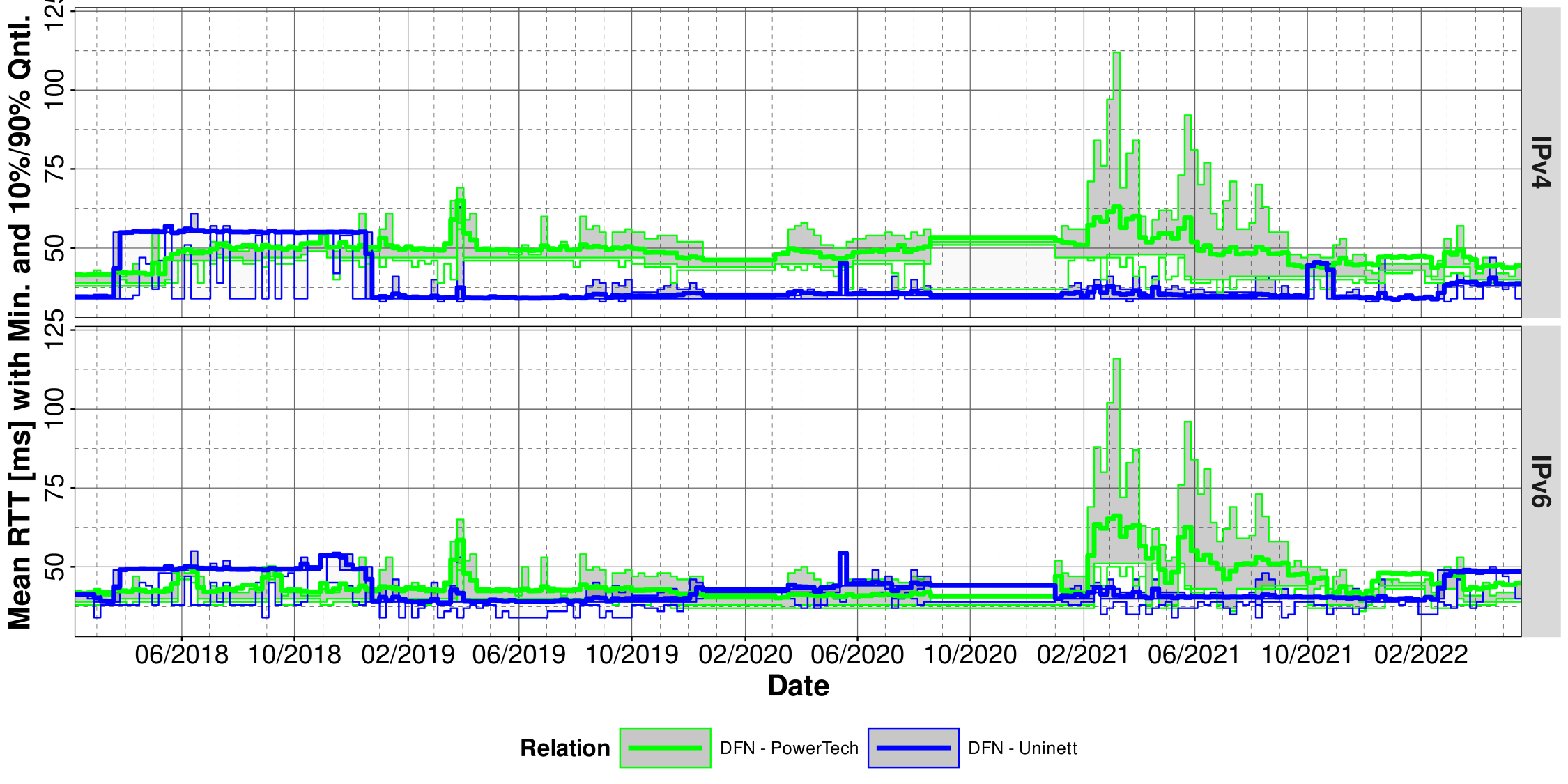}
\end{center}
\caption{RTT Time Series for Essen, DE to Trondheim, NO.}
\label{fig:TimeSeries-UDE-NTNU}
\end{figure*}

\begin{figure*}
\begin{center}
\includegraphics[width=1.00\textwidth]{%
   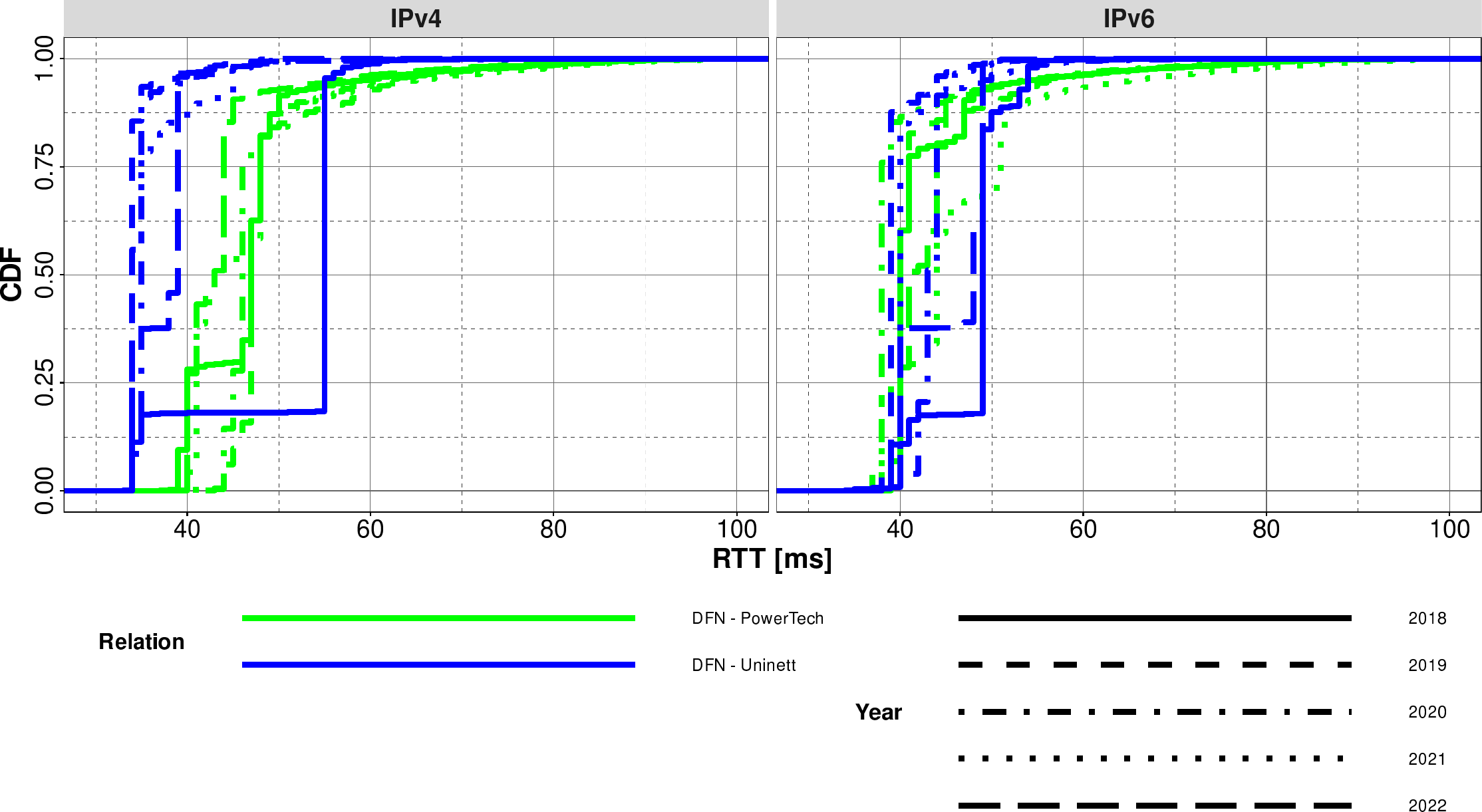}
\end{center}
\caption{RTT CDF for Essen, DE to Trondheim, NO.}
\label{fig:CDF-UDE-NTNU}
\end{figure*}
To examine the scenario of non-neighboring countries located on the same continent, we analyze the RTTs between Essen, \worldflag[width=6pt,length=10pt,stretch=1]{DE}DE and Trondheim, \worldflag[width=6pt,length=10pt,stretch=1]{NO}NO. The site in Essen is connected to DFN (\foreignlanguage{german}{Deutsches Forschungsnetz}, the German national research and education network), which supports IPv4 and IPv6.
Figure~\ref{fig:TimeSeries-UDE-NTNU} shows the RTT time series with mean, as well as 10\%- and 90\%-quantiles. Figure~\ref{fig:CDF-UDE-NTNU} presents the corresponding average RTT CDF. Both figures are split between IPv4 and IPv6.

First, the general observation also seen for the neighboring countries scenario of Subsubsection~\ref{subsub:Ping-KAU-NTNU} can be confirmed again: the connectivity to Uninett is also relatively stable, with average as well as 10\%- and 90\%-quantiles being similar. The only exception is around~2018, where the RTTs were about 20~ms to 30~ms higher, with a similar effect for IPv4 and IPv6. Again, PowerTech shows more variation, with the peaks mostly identical for IPv4 and IPv6. However, an exciting difference is visible here: \textit{IPv6 has a lower RTT than IPv4} for PowerTech. Compared to Uninett, the PowerTech IPv6 RTTs are significantly lower for most of the time, despite of using a higher-delay ADSL connection instead of high-speed, low-delay fibre. The primary reason for such observations lies within the ISP interconnections, which we examine in detail in Subsection~\ref{sub:AS-Level-Findings} and Subsubsection~\ref{subsub:Traceroute-UDE-NTNU}.
Since both protocols are routed independently, one protocol may have a shorter route than the other, leading to a lower RTT for one of them.
Next, we take a look at the RTTs of an inter-continental scenario.

\subsubsection{Scenario: Inter-Continental}
\label{subsub:Ping-HU-NTNU}

\begin{figure*}
\begin{center}
\includegraphics[width=1.00\textwidth]{%
   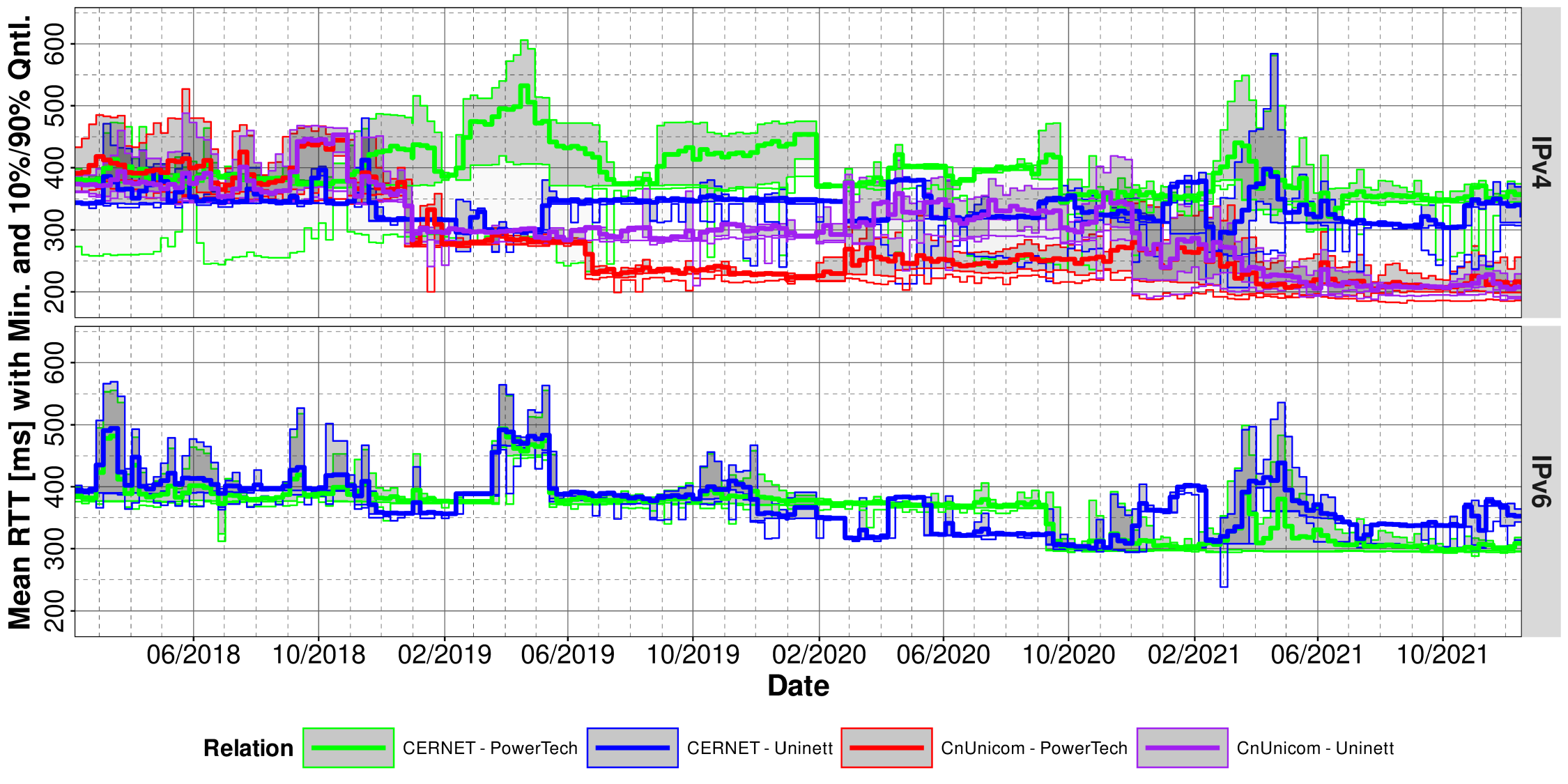}
\end{center}
\caption{RTT Time Series for Haikou, CN to Trondheim, NO.}
\label{fig:TimeSeries-HU-NTNU}
\end{figure*}

\begin{figure*}
\begin{center}
\includegraphics[width=1.00\textwidth]{%
   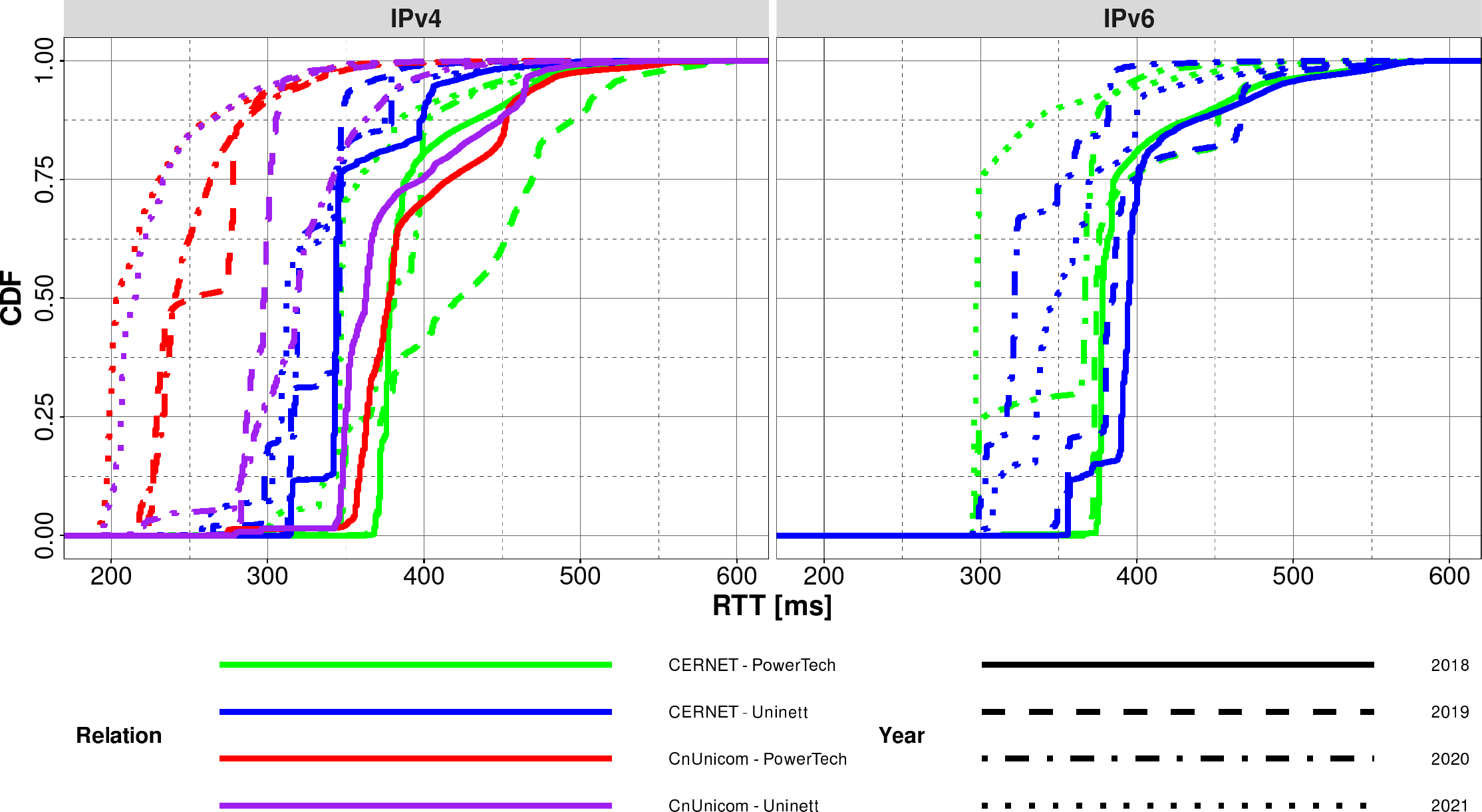}
\end{center}
\caption{RTT CDF for Haikou, CN to Trondheim, NO.}
\label{fig:CDF-HU-NTNU}
\end{figure*}
We analyze the RTTs between Haikou, \worldflag[width=6pt,length=10pt,stretch=1]{CN}CN and Trondheim, \worldflag[width=6pt,length=10pt,stretch=1]{NO}NO for the inter-continental scenario. The site in Haikou is connected to two ISPs. They are CERNET (China Education and Research Network) and China Unicom (a business-grade fibre network). Both ISPs offer IPv4 as well as IPv6 connectivity. Unfortunately, \texttt{ICMP Echo Reply} over IPv6 is blocked on China Unicom, preventing IPv6 RTT measurements here. Also, there is only data until December~2021 available, since from January~2022 the firewall requirements of the site host in China have become stricter.
Figure~\ref{fig:TimeSeries-HU-NTNU} presents the time series with average 10\%- and 90\%-quantiles, while Figure~\ref{fig:CDF-HU-NTNU} shows the mean RTT CDF. The four relations are depicted in different colors: CERNET$\rightarrow$PowerTech (in green), CERNET$\rightarrow$Uninett (in blue), China Unicom$\rightarrow$Power Tech (in red) and China Unicom$\rightarrow$Uninett (in purple).

Depending on the choice of ISP on each side, there is a significantly different behavior. Interestingly, in~2021, PowerTech with its high-delay ADSL connection had the smallest RTT of all 4~combinations over China Unicom (dotted, red). In contrast, compared to CERNET, the connection to PowerTech had the highest RTT (dotted, green). The average RTT, and the 10\%- and 90\%-quantiles, have also significantly changed during the 4~years of observation, with a tendency to decrease the mean RTT and variation over time. The same can be observed for IPv6 (via CERNET only).

As a summary of the observations, it can be seen that the data packet route is not static and fixed. It leads to the following question: \textit{What are the reasons?} and \textit{What are the implications for a network user}?

\subsection{Routing Findings}
\label{sub:AS-Level-Findings}
To further examine the reasons for the mentioned RTT behavior observed in the previous subsection, we analyze the observed routing from the \noun{HiPerConTracer} Traceroute results and map links to ASs (i.e., ISPs) and countries. From the results, we can conclude:
\begin{enumerate}
 \item Data packet routing between two fixed points can involve third-party countries;
 \item Traffic may take unexpected routes;
 \item Routes change over time between two fixed points;
 \item IPv4 and IPv6 may take very different routes.
\end{enumerate}

\subsubsection{Scenario: Between Neighboring Countries}
\label{subsub:Traceroute-KAU-NTNU}

\begin{figure*}
\begin{center}
\includegraphics[width=1.00\textwidth]{%
   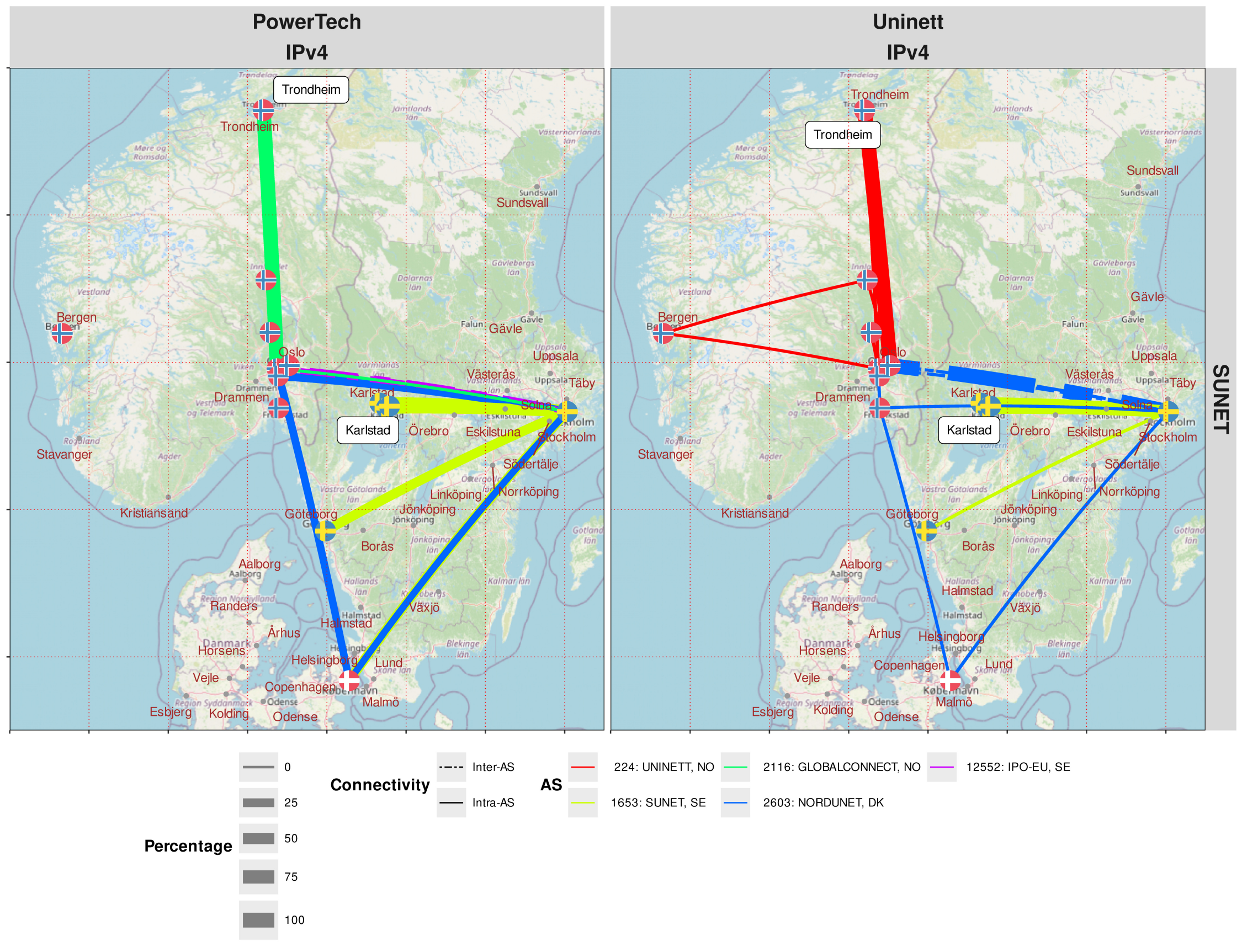}
\end{center}
\caption{Observed routes and ASs for data packets from Karlstad, SE to Trondheim, NO.}
\label{fig:ASMap-KAU-NTNU}
\end{figure*}
First, we look at the routing scenario of neighboring countries (see Subsubsection~\ref{subsub:Ping-KAU-NTNU}) from Karlstad, \worldflag[width=6pt,length=10pt,stretch=1]{SE}SE, to Trondheim, \worldflag[width=6pt,length=10pt,stretch=1]{NO}NO. The site in Karlstad only has IPv4 connectivity via its only ISP, SUNET, while the site in Trondheim is connected to Uninett and PowerTech. Figure~\ref{fig:ASMap-KAU-NTNU} presents the observed routers, their geo-location (see Subsection~\ref{sub:Testbed-Setup}), and the AS mapping of the observed links. It is split between PowerTech (left) and Uninett (right) as incoming ISPs of the site in Trondheim. Dashed lines visually represent intra-AS links (between different ASs), while solid lines represent intra-AS links (within the same AS). The color of the link corresponds to the \textit{source} AS of a link. The line thickness shows the relative number of observations (from 0.1~\% to 100~\%) in all \noun{HiPerConTracer} Traceroute runs. Links observed in less than 0.1~\%~of the runs are filtered to hide extremely rarely seen links, which are the temporary effect of route changes. \noun{HiPerConTracer} Traceroute relies on the ICMP responses from the routers. If a router does not respond, its address and associated geo-location of its links remain unknown and therefore cannot be displayed on the map or taken into account for percentage computations. A router may completely block ICMP traffic or limit the ICMP response rate due to the applied security rules.

\input{Measurements/TEX/Traceroute-HUUDEKAU-NTNU-KAU-NTNU-AS-P0.1.tex}
\input{Measurements/TEX/Traceroute-HUUDEKAU-NTNU-KAU-NTNU-Country-P0.1.tex}
Table~\ref{tab:AS-KAU-NTNU} provides the percentages between AS, aggregating routers with the ASs they belong to, showing the influence of ISPs and their interconnections. The percentage provides the fraction of \noun{HiPerConTracer} Traceroute runs in which a corresponding inter-AS forwarding had been observed for the given combination of outgoing ISP (SUNET in Karlstad), incoming ISP (Uninett or PowerTech in Trondheim), and IP version (IPv4 only).

It is important to note that percentages cannot contain observations in which all routers in an AS have not responded\footnote{E.g.,\ due to packet loss by congestion or ICMP response rate limitation.} for a Traceroute run, i.e., the actual percentages may be slightly higher. Similarly, Table~\ref{tab:Country-KAU-NTNU} provides an individual country-level aggregation with cross-border data traffic percentages. The table shows the mean RTT and 10~\%/90~\%-quantiles of the RTT between the source in Karlstad and the routers of the destination AS. The RTTs are provided as a rough estimate of \textit{how far away} (referring to transmission time) the corresponding AS is from the source.

All traffic must leave the site via SUNET (AS~1653 in orange) via IPv4, as the Karlstad site only supports IPv4 connectivity. The Trondheim site is connected to two ISPs: Uninett (AS~224 in red) and PowerTech (a subsidiary of GlobalConnect, AS~2116 in green). Therefore, the incoming traffic is split between Uninett and GlobalConnect. Although research networks (Uninett in Norway and SUNET in Sweden) are mainly connected within their corresponding country, possible connections between Karlstad and Trondheim can involve inter-country links from NorduNet (connects national research networks of the Nordic countries; AS~2603 in blue) as well as GlobalConnect and IPO-EU (AS~12552; in purple).

Between Karlstad and Trondheim, multiple routes are possible, for example, via Copenhagen or Stockholm. Traffic can take significant detours compared to the geographically shortest route. NorduNet connects Sweden and Norway via Denmark. It implies that data packet traffic between the neighboring countries Sweden and Norway may go through the third-party country Denmark.

For SUNET$\rightarrow$PowerTech, approximately half of the observations (48.35\%; see Table~\ref{tab:Country-KAU-NTNU}) have a border crossing SE$\rightarrow$DK (via NorduNet; see Table~\ref{tab:AS-KAU-NTNU}), while the other half (50.69~\%) have a direct transfer SE$\rightarrow$NO. Note that traffic may re-enter SE (then via NorduNet). SUNET$\rightarrow$Uninett almost always takes the direct route (99.07~\%\footnote{Again, note that lost or not sent ICMP responses cannot be counted.}) via IPO-EU or NorduNet. Only in sporadic cases, Uninett traffic is routed via DK (0.56~\%) and may either go from there to NO (0.25~\%) or enter SE again (0.30~\%; via NorduNet). So, traffic may be routed through or even take a ``detour'' via another country. In this scenario, all countries belong to the European Economic Area~(EEA) and apply similar data communication rules. It is also worth noting that the data packet movement is influenced by many factors implemented by ISPs: the research network providers use their own, relatively static infrastructures (SUNET, Uninett, NorduNet), while commercial ISPs like PowerTech try forwarding to other ISPs to minimize costs. The latter leads to more frequent changes over time, as observed for PowerTech. These changes particularly affect RTTs (shorter or longer routes; for RTTs, see Figure~\ref{fig:CDF-KAU-NTNU} and Subsubsection~\ref{subsub:Traceroute-KAU-NTNU}).

\input{Measurements/TEX/Traceroute-HUUDEKAU-NTNU-KAU-NTNU-Hops.tex}
To provide an overview of the path lengths observed during the 5~years of observations, Table~\ref{tab:Hops-KAU-NTNU} displays the hop count statistics: absolute minimum number of hops, 10\%-/90\%-quantiles (to show the typical range), as well as mean and median. In this small scenario, both ISPs had very similar mean and median hop counts. However, the 10\%-/90\%-quantiles range for PowerTech is larger (13--16) than that for Uninett (14--15). In other words, the commercial ISP exhibits a greater variation in path length.

\subsubsection{Scenario: Intra-Continental}
\label{subsub:Traceroute-UDE-NTNU}

\begin{figure*}
\begin{center}
\includegraphics[width=0.9575\textwidth]{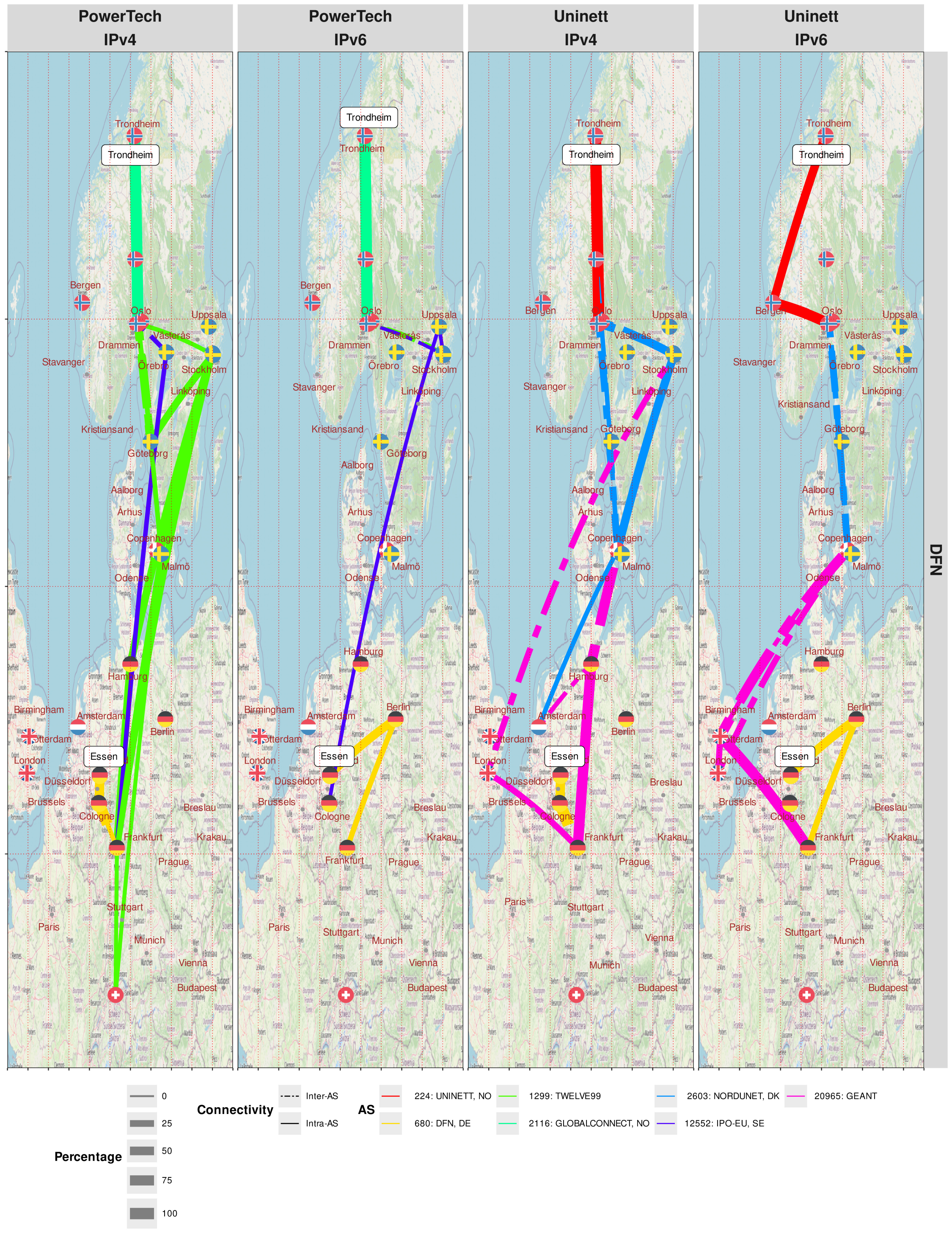}
\end{center}
\caption{Observed routes and ASs for both IPv4 and IPv6 data packets from Essen, DE to Trondheim, NO.}
\label{fig:ASMap-UDE-NTNU}
\end{figure*}
Next, we will look at intra-continental communication between two non-neighboring countries (see Subsubsection~\ref{subsub:Ping-UDE-NTNU}). Figure~\ref{fig:ASMap-UDE-NTNU} presents the data traffic from Essen, \worldflag[width=6pt,length=10pt,stretch=1]{DE}DE to Trondheim, \worldflag[width=6pt,length=10pt,stretch=1]{NO}NO (only percentages $\ge$2.5~\% are shown). Unlike the scenario of neighboring countries in Subsubsection~\ref{subsub:Traceroute-KAU-NTNU}, where the countries share a border, this scenario necessarily involves third-party countries. That is, although the geographic distance is just around 1,400~km, a significantly larger number of countries (namely CH, DK, SE, GB, NL)\footnote{Denmark~(\worldflag[width=6pt,length=10pt,stretch=1]{DK}DK), Great Britain~(\worldflag[width=6pt,length=10pt,stretch=1]{GB}GB), Netherlands~(\worldflag[width=6pt,length=10pt,stretch=1]{NL}NL), Switzerland~(\worldflag[width=6pt,length=10pt,stretch=1]{CH}CH).} is seen here. Data traffic may even leave the EEA, and get routed via CH (DFN$\rightarrow$Uninett via IPv4; DFN$\rightarrow$PowerTech via IPv4) and GB (DFN$\rightarrow$Uninett via IPv4 and IPv6).

\input{Measurements/TEX/Traceroute-HUUDEKAU-NTNU-UDE-NTNU-Country-P2.5.tex}

There are many routes between DE and NO, and Table~\ref{tab:Country-UDE-NTNU} presents the percentages of border crossing traffic for IPv4 and IPv6. Routes change over time, leading to smaller percentages compared to the scenario of neighboring countries in Subsection~\ref{subsub:Traceroute-KAU-NTNU}.
As expected, there are again differences between the involved ISPs.
While PowerTech takes a straightforward route through Denmark and Sweden (DE$\rightarrow$DK$\rightarrow$SE$\rightarrow$NO), Uninett routes via GB (DE$\rightarrow$GB$\rightarrow$DK$\rightarrow$SE$\rightarrow$NO). There may be hidden links, particularly for IPv6 leaving DE: only 3.24~\% (PowerTech: DE$\rightarrow$SE) compared to 99.86~\% (Uninett: DE$\rightarrow$GB) of the IPv6 observations show a country crossing from DE. In particular, the longer distances of the IPv6 Uninett paths compared to the more direct PowerTech paths then lead to the increase in RTT observed in Subsection~\ref{subsub:Ping-UDE-NTNU} (see also Figure~\ref{fig:CDF-UDE-NTNU}).

Furthermore, using the country RTTs of our \noun{HiPerConTracer} Traceroute runs from Table~\ref{tab:Country-UDE-NTNU}, we can identify and filter out some false positives: An initially seen ``detour'' SE$\rightarrow$US\footnote{United States (\worldflag[width=6pt,length=10pt,stretch=1]{US}US).}$\rightarrow$SE, with a minimum round-trip distance of approximately 12,500~km, would add \textit{at least} 41~ms for speed $c_0$\footnote{Light speed in a vacuum: $c_0$=299792.458~$\frac{\mathrm{km}}{s}$.}, which is not the case. However, we observed routes through the US with realistic RTTs on a few rare occasions, probably on temporary route changes (0.14~\%, not shown in Table~\ref{tab:Country-UDE-NTNU}).
Nevertheless, with CH, GB (and possibly even the US), traffic between EEA countries may be routed via non-EEA countries.

\input{Measurements/TEX/Traceroute-HUUDEKAU-NTNU-UDE-NTNU-AS-P2.5.tex}
The observation percentages of inter-AS connectivity are shown in Table~\ref{tab:AS-UDE-NTNU}: Here, it is possible to see that the data between the two research networks (i.e.\ DFN$\rightarrow$Uninett) usually uses research network providers for the exchange, i.e.\ NorduNet (AS~2603; in light-blue in Figure~\ref{fig:ASMap-UDE-NTNU}) and GÉANT (AS~20965; in pink). On the other hand, the connection to the commercial ISP PowerTech goes via IPO-EU (AS~12552; in dark blue) and Twelve99 (AS~1299; in green). Interestingly, although the same ASs are involved for both versions of the protocol, the routing significantly differs between IPv4 and IPv6. That is, it may \textit{not} be assumed that IPv4 and IPv6 provide similar performance regarding end-user perceived experience, even for scenarios within the same continent.

\input{Measurements/TEX/Traceroute-HUUDEKAU-NTNU-UDE-NTNU-Hops.tex}
Furthermore, Table~\ref{tab:Hops-UDE-NTNU} shows the hop count statistics for the 5~years of observations. A comparison of the research network ISP Uninett and the commercial ISP PowerTech shows that PowerTech requires fewer hops on the mean, and the median requires fewer hops (13 vs.\ 17 for median). In other words, the PowerTech routes are shorter. This applies to both IPv4 and IPv6, although the routes of both protocols can differ significantly, as shown in Figure~\ref{fig:ASMap-UDE-NTNU}.

\subsubsection{Scenario: Inter-Continental}
\label{subsub:Traceroute-HU-NTNU}

\begin{figure*}
\begin{center}
\includegraphics[width=1.00\textwidth]{%
   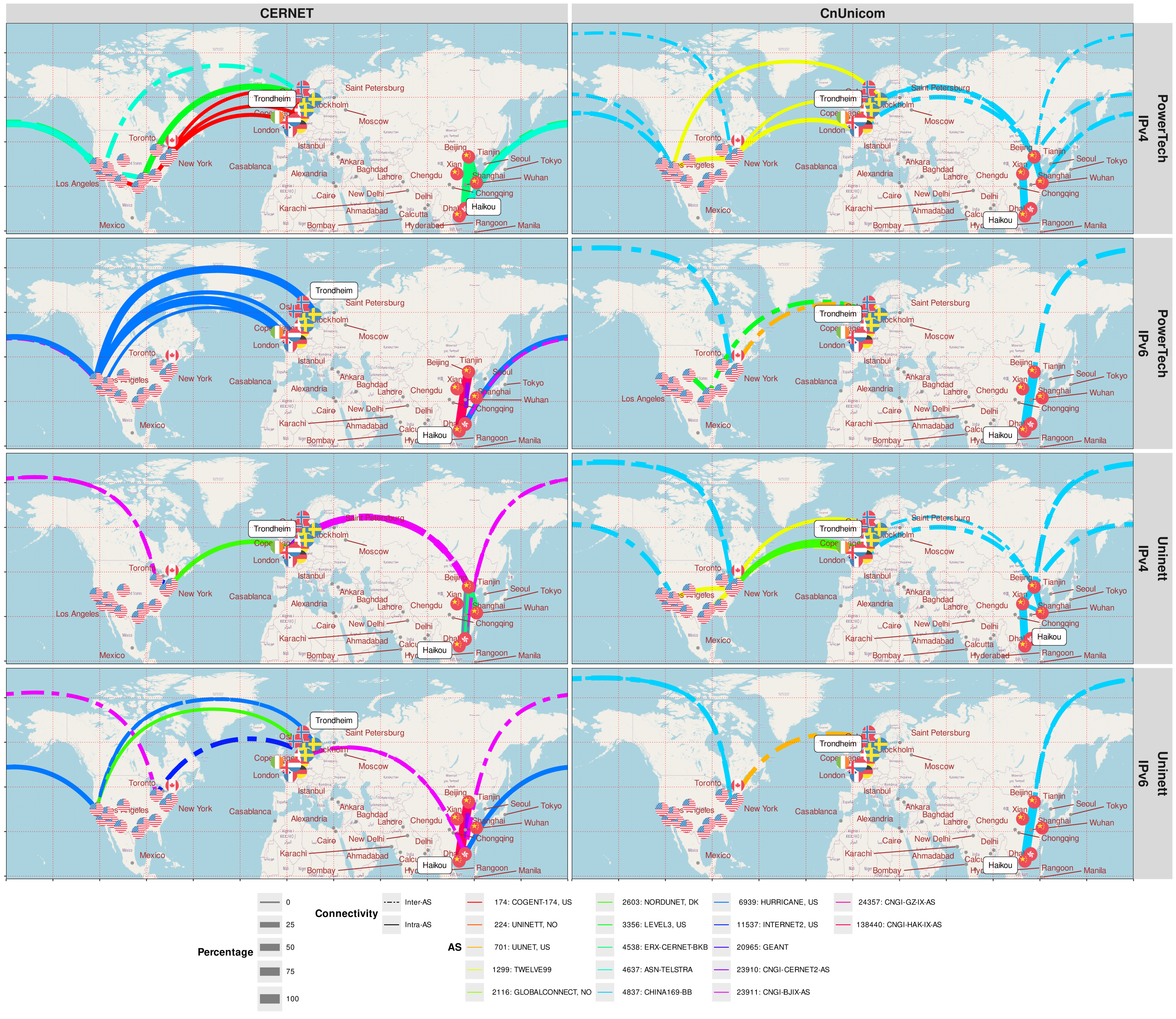}
\end{center}
\caption{Observed routes and ASs for both IPv4 and IPv6 data packets from Haikou, CN to Trondheim, NO.}
\label{fig:ASMap-HU-NTNU}
\end{figure*}
Finally, Figure~\ref{fig:ASMap-HU-NTNU} presents the results visualisation for the inter-continental setup (see also Subsubsection~\ref{subsub:Ping-HU-NTNU}) for the traffic from Haikou, CN to Trondheim, NO (again, only percentages $\ge$2.5~\% are shown). Similar to the previous two scenarios, such communication involves various regions and third-party countries, such as BE, CA, DE, DK, FR, GB, HK, IE, NL, SE, SG, US\footnote{Belgium~(\worldflag[width=6pt,length=10pt,stretch=1]{BE}BE), Canada~(\worldflag[width=6pt,length=10pt,stretch=1]{CA}CA), France~(\worldflag[width=6pt,length=10pt,stretch=1]{FR}FR), Hong Kong~(\worldflag[width=6pt,length=10pt,stretch=1]{CN}HK; special administrative region of China), Ireland~(\worldflag[width=6pt,length=10pt,stretch=1]{IE}IE), Singapore~(\worldflag[width=6pt,length=10pt,stretch=1]{SG}SG).}, while the geographically shortest distance is just around 8,500~km (Asia to Europe via Russia). We can see that most of the observed routes take the eastern direction from China via different trans-Pacific and trans-Atlantic cables, because the US is well-connected to Asia and Europe.

\input{Measurements/TEX/Traceroute-HUUDEKAU-NTNU-HU-NTNU-Country-P15.tex}
Table~\ref{tab:Country-HU-NTNU} shows the observed percentages of IPv4 and IPv6 data traffic for the Haikou, CN to Trondheim, NO scenario. Due to space limitations, only the border crossing percentages of $\ge$15~\% are displayed here. The table reflects the results of Figure~\ref{fig:ASMap-HU-NTNU}, with many different routing possibilities. Most traffic uses routes eastwards via the US. The only exception is CERNET$\rightarrow$Uninett over IPv4, where 74.46~\% of the Traceroute runs show a direct link CN$\rightarrow$GB. For IPv6, the same relation is only seen in 18.18~\% of the runs. When comparing the two ISPs, i.e., the research network ISP CERNET and the commercial ISP China Unicom, it can be seen that there is much more variation for CERNET. Routing from China Unicom mainly involves only US, SE, and DK as third-party countries, while other countries can be seen when routing from CERNET. That is, the two ISPs use very different routing strategies.

\input{Measurements/TEX/Traceroute-HUUDEKAU-NTNU-HU-NTNU-AS-P10.tex}
The corresponding percentages of ASs are shown in Table~\ref{tab:AS-HU-NTNU} (only percentages $\ge$10~\%, due to space limitations). Here, observations from country border crossings can be confirmed: while routing from China Unicom only involves routing via UUNET (AS~701; in orange in Figure~\ref{fig:ASMap-HU-NTNU}), Twelve99 (AS~1299; in yellow), NorduNet (AS~2603; in green) and Level3 (AS~3356; in light-green), various more ASs can be seen for the routing from CERNET. During the four years of observations, multiple routes have been observed between the two sites, involving 20~different ASs and many third-party countries. The observed routes are probably the result of economic decisions made by the ISPs involved, which change over time~\cite{FTC2021}. These changes have an apparent effect on the RTTs (see Subsection~\ref{sub:Connectivity-Findings}).

\input{Measurements/TEX/Traceroute-HUUDEKAU-NTNU-HU-NTNU-Hops.tex}
Finally, Table~\ref{tab:Hops-HU-NTNU} shows the hop count statistics for the inter-continental scenario: the results differ significantly depending on the choice of source and destination ISP, and IP protocol. Notable is the absolute minimum number of required hops: only 13 for China Unicom to Uninett over IPv4 and 20 for CERNET to PowerTech over IPv6. Furthermore, there is a wide range between 10\%- and 90\%-quantiles in many cases (e.g.\ 21--31 for CERNET to PowerTech over IPv4), with only China Unicom to Uninett over IPv4 the exception (16--18, i.e.\ a relatively stable path length over time).

\section{Discussions, Limitations and Applicability}
Although it is true that the IP datagram service is connectionless, transport protocols (like TCP, Stream Control Transmission Protocol~(SCTP)~\cite{RFC9260}, Datagram Congestion Control Protocol (DCCP)~\cite{RFC4340}, and application protocols on top of UDP) make an implicit assumption that ``generally'' packets of the same connection also take the same route, or at least there is ``usually'' no reordering. If there is reordering, the typical loss-based congestion control (of TCP, SCTP, DCCP, or UDP-based application protocols) sees packet losses due to detected gaps in the packet sequence, therefore assumes congestion and reduces the output rate to mitigate this congestion. This leads to poor network packet transmission performance.

We found that path selection is a complex process, and the data path is not fixed. We observed a significant deviation from the shortest path distance. The shortest land distance is sometimes only counted when routing data packets between two fixed points. User packets take different routes that cover many countries and detours, even between two fixed points. We also found that commercial ISPs prefer commercial-grade ISPs, whereas research ISPs prefer research network ISPs. Such findings help us to answer our first research sub-question, that is, \textit{How ISP allocate the data traffic route between two fixed points?}

Next, we found that network latency can be changed by changing the ISP and IP protocols from IPv4 to IPv6. These findings should be considered in general because six large ISPs that cover research networks, consumer-grade ADSL lines, and commercial-grade commercial fibers are used to collect traces. RTT changes caused by ISPs can affect network performance and user experience. Fluctuations in the RTT may lead to unpredictable latency, which can reduce the user's perceived experience about an application. TCP is based on RTT measurements for congestion control and flow control. Even for the same ISP, different RTTs can exist for IPv4 and IPv6. One or another protocol may perform better. For better performance, the preferred protocol depends on the sites and ISPs and may change over time. Therefore, we also answered our second research sub-question, that is, \textit{How does network performance get affected while data are in transit?}

\subsection{Limitations}
We have also observed the limitations of our research work. They are the following:
\begin{enumerate}
    \item \textit{Traceroute and others:} As mentioned previously, Traceroute has some limitations; however, it is less intrusive to identify routing problems. Even if some ISPs share routing information, obtaining data from all or at least most ISPs is doubtful. The popular BGP data are not intended to infer AS-level mappings. By definition, BGP was not designed with an AS-level topology discovery feature~\cite[Subsubsection~I.A.4]{Roughan201110}. The BGP route collector infrastructure provides complete AS-level connectivity information for a very small number of ASes~\cite{Gregori2012incompleteness}. In addition, BGP suffers from misconfiguration errors of routes that can propagate across the backbone of the Internet~\cite{Mahajan2022}. Furthermore, the RIPE Atlas data do not accurately represent global Internet connectivity~\cite{Singh2018characterizing}.

   \item \textit{Data Limitations} For analysis, we use four vantage points. The data presents one direction of the routing rather than both, due to firewall limitations.

    \item \textit{Infrastructure Management:} We had to continue running the \noun{NorNet} infrastructure for over 5~years, spanning multiple national and international locations. Due to that, negotiations were made with various entities, primarily the measurement site hosts (refer to~\ref{sec:measure_infra} for more).

   \item \textit{Accuracy:} Mapping the small countries and short geographic distances (a few hundred kilometres) in Europe was challenging. To counter the issue, we perform approximations using RTTs by HLOC~\cite{HLOC2017} during traffic analysis.
\end{enumerate}

\subsection{Applicability}
Generally, in packet-switched networks, deterministic data packet transmission paths are helpful to \textit{i)} ensure consistent and predictable network performance; \textit{ii)} achieve better service provision; \textit{iii)} optimize network resource utilization, and improve traffic management. Dynamic routing protocols offer flexibility and adaptability, but fixed routes offer stability and better physical network security. Deterministic paths would be nice for better service, but the Internet does not yet provide them.

Latency is an indicator of network health. It can play an important role in supporting latency-sensitive applications. Indirectly, it may also play a role in workload balancing. Following this metric, a network performance engineer can opt for alternate paths with lower latency. Our findings can help network performance engineers to improve traffic engineering. Although the exact reason for a route change may remain unknown (due to non-disclosure of infrastructure level information from ISPs), its effect on the RTT is very important from an end-user perspective.



\section{Conclusions and Future Work}
Packets taking unexpected detours and changes in routing impact higher-level protocols. The aim of this paper is to illustrate the impacts of long-and short-term network effects based on a measurement analysis over five years with three different real-world data communication scenarios (such as neighboring countries, intra-continental, inter-continental). The proposed method is based on measurements on \emph{own} end-systems without the need for data from the underlying ISPs. For the analysis, we developed an open-source measurement framework \noun{HiPerConTracer} for Ping and Traceroute runs.

As part of our ongoing work, we will further analyze \noun{HiPerConTracer} Ping and Traceroute data in more detail, possibly applying machine learning and combining it with publicly available BGP data to learn more about routing changes. Furthermore, we would like to further improve the geo-location of addresses and perform a detailed analysis of transitory phase changes along the route. That is, how frequent are the changes, for how long and how much do these changes affect the RTTs, and was there an observable reason for the change? In addition, ongoing and future work will include analyzing the performance impact of higher-level protocols (e.g.\ TCP, SCTP) on congestion-control-related throughput estimates (based on collected RTT data) and packet loss and jitter. At the end of this project, configuration guidelines for network performance are provided.

\bibliographystyle{elsarticle-num}
\bibliography{ref}

\end{document}

%% file: Measurements/TEX/Traceroute-HUUDEKAU-NTNU-KAU-NTNU-AS-P0.1.tex
\begin{table*}
\centering
\caption{Inter-AS Observations Percentage ($\ge$0.1) and Mean RTT as well as 10\%-/90\%-Quantiles RTT (in ms) for Karlstad, SE $\rightarrow$ Trondheim, NO} 
\label{tab:AS-KAU-NTNU}
\begin{adjustbox}{max width=\textwidth}
\small
\begin{tabular}{|c|c|c|c|c|c|c|c|c|}
  \hline
{\textbf{IP}} & {\textbf{From ISP}} & {\textbf{To ISP}} & {\textbf{From}} & {\textbf{To}} & {\textbf{Mean}} & {\textbf{Q$_{\mathrm{10\%}}$}} & {\textbf{Q$_{\mathrm{90\%}}$}} & {\textbf{\%}} \\ 
  \hline
\hline
IPv4 & SUNET & Uninett &   1653: SUNET &   2603: NORDUNET & 13.80 & 11.76 & 14.32 & 99.41 \\ 
   \hline
IPv4 & SUNET & Uninett &   2603: NORDUNET &    224: UNINETT & 26.90 & 24.50 & 27.62 & 99.32 \\ 
   \hline
\end{tabular}
\end{adjustbox}
\end{table*}

%% file: Measurements/TEX/Traceroute-HUUDEKAU-NTNU-KAU-NTNU-Country-P0.1.tex
\begin{table*}
\centering
\caption{Inter-Country Observations Percentage ($\ge$0.1) and Mean RTT as well as 10\%-/90\%-Quantiles RTT (in ms) for Karlstad, SE $\rightarrow$ Trondheim, NO} 
\label{tab:Country-KAU-NTNU}
\begin{adjustbox}{max width=\textwidth}
\small
\begin{tabular}{|c|c|c|c|c|c|c|c|c|}
  \hline
{\textbf{IP}} & {\textbf{From ISP}} & {\textbf{To ISP}} & {\textbf{From}} & {\textbf{To}} & {\textbf{Mean}} & {\textbf{Q$_{\mathrm{10\%}}$}} & {\textbf{Q$_{\mathrm{90\%}}$}} & {\textbf{\%}} \\ 
  \hline
\hline
IPv4 & SUNET & Uninett & \worldflag[width=6pt,length=10pt,stretch=1]{SE}SE & \worldflag[width=6pt,length=10pt,stretch=1]{NO}NO & 26.90 & 24.50 & 27.62 & 99.07 \\ 
   \hline
IPv4 & SUNET & Uninett & \worldflag[width=6pt,length=10pt,stretch=1]{SE}SE & \worldflag[width=6pt,length=10pt,stretch=1]{DK}DK & 9.13 & 6.75 & 11.25 & 0.56 \\ 
   \hline
IPv4 & SUNET & Uninett & \worldflag[width=6pt,length=10pt,stretch=1]{DK}DK & \worldflag[width=6pt,length=10pt,stretch=1]{SE}SE & 17.78 & 16.26 & 18.86 & 0.30 \\ 
   \hline
IPv4 & SUNET & Uninett & \worldflag[width=6pt,length=10pt,stretch=1]{DK}DK & \worldflag[width=6pt,length=10pt,stretch=1]{NO}NO & 16.81 & 13.26 & 18.95 & 0.25 \\ 
   \hline
\end{tabular}
\end{adjustbox}
\end{table*}

%% file: Measurements/TEX/Traceroute-HUUDEKAU-NTNU-KAU-NTNU-Hops.tex
\begin{table*}
\centering
\caption{Hop count statistics for Karlstad, SE $\rightarrow$ Trondheim, NO} 
\label{tab:Hops-KAU-NTNU}
\begin{adjustbox}{max width=\textwidth}
\small
\begin{tabular}{|c|c|c|c|c|c|c|c|}
  \hline
{\textbf{IP}} & {\textbf{From ISP}} & {\textbf{To ISP}} & {\textbf{Min}} & {\textbf{Q$_{\mathrm{10\%}}$}} & {\textbf{Mean}} & {\textbf{Median}} & {\textbf{Q$_{\mathrm{90\%}}$}} \\ 
  \hline
\hline
IPv4 & SUNET & Uninett &  14 & 14.00 & 14.66 & 15.00 & 15.00 \\ 
   \hline
\end{tabular}
\end{adjustbox}
\end{table*}

%% file: Measurements/TEX/Traceroute-HUUDEKAU-NTNU-UDE-NTNU-Country-P2.5.tex
\begin{table*}
\centering
\caption{Inter-Country Observations Percentage ($\ge$2.5) and Mean RTT as well as 10\%-/90\%-Quantiles RTT (in ms) for Essen, DE $\rightarrow$ Trondheim, NO} 
\label{tab:Country-UDE-NTNU}
\begin{adjustbox}{max width=\textwidth}
\small
\begin{tabular}{|c|c|c|c|c|c|c|c|c|}
  \hline
{\textbf{IP}} & {\textbf{From ISP}} & {\textbf{To ISP}} & {\textbf{From}} & {\textbf{To}} & {\textbf{Mean}} & {\textbf{Q$_{\mathrm{10\%}}$}} & {\textbf{Q$_{\mathrm{90\%}}$}} & {\textbf{\%}} \\ 
  \hline
\hline
IPv4 & DFN & PowerTech & \worldflag[width=6pt,length=10pt,stretch=1]{SE}SE & \worldflag[width=6pt,length=10pt,stretch=1]{NO}NO & 31.15 & 26.13 & 34.82 & 99.53 \\ 
   \hline
IPv4 & DFN & PowerTech & \worldflag[width=6pt,length=10pt,stretch=1]{DE}DE & \worldflag[width=6pt,length=10pt,stretch=1]{SE}SE & 21.53 & 17.75 & 26.79 & 92.84 \\ 
   \hline
IPv4 & DFN & PowerTech & \worldflag[width=6pt,length=10pt,stretch=1]{NO}NO & \worldflag[width=6pt,length=10pt,stretch=1]{SE}SE & 33.80 & 32.43 & 37.77 & 9.60 \\ 
   \hline
IPv4 & DFN & PowerTech & \worldflag[width=6pt,length=10pt,stretch=1]{DE}DE & \worldflag[width=6pt,length=10pt,stretch=1]{CH}CH & 6.53 & 6.25 & 6.71 & 4.52 \\ 
   \hline
IPv4 & DFN & PowerTech & \worldflag[width=6pt,length=10pt,stretch=1]{CH}CH & \worldflag[width=6pt,length=10pt,stretch=1]{SE}SE & 24.69 & 12.59 & 62.65 & 4.51 \\ 
   \hline
IPv4 & DFN & Uninett & \worldflag[width=6pt,length=10pt,stretch=1]{DE}DE & \worldflag[width=6pt,length=10pt,stretch=1]{DK}DK & 19.71 & 18.36 & 22.11 & 73.09 \\ 
   \hline
IPv4 & DFN & Uninett & \worldflag[width=6pt,length=10pt,stretch=1]{SE}SE & \worldflag[width=6pt,length=10pt,stretch=1]{NO}NO & 33.95 & 26.36 & 47.42 & 63.28 \\ 
   \hline
IPv4 & DFN & Uninett & \worldflag[width=6pt,length=10pt,stretch=1]{DK}DK & \worldflag[width=6pt,length=10pt,stretch=1]{SE}SE & 28.31 & 26.68 & 29.37 & 44.21 \\ 
   \hline
IPv4 & DFN & Uninett & \worldflag[width=6pt,length=10pt,stretch=1]{DK}DK & \worldflag[width=6pt,length=10pt,stretch=1]{NO}NO & 30.73 & 26.61 & 43.49 & 36.49 \\ 
   \hline
IPv4 & DFN & Uninett & \worldflag[width=6pt,length=10pt,stretch=1]{DE}DE & \worldflag[width=6pt,length=10pt,stretch=1]{GB}GB & 13.35 & 12.46 & 13.70 & 19.25 \\ 
   \hline
IPv4 & DFN & Uninett & \worldflag[width=6pt,length=10pt,stretch=1]{GB}GB & \worldflag[width=6pt,length=10pt,stretch=1]{SE}SE & 40.33 & 39.57 & 40.61 & 19.22 \\ 
   \hline
IPv4 & DFN & Uninett & \worldflag[width=6pt,length=10pt,stretch=1]{DE}DE & \worldflag[width=6pt,length=10pt,stretch=1]{NL}NL & 21.17 & 19.95 & 23.16 & 7.43 \\ 
   \hline
IPv4 & DFN & Uninett & \worldflag[width=6pt,length=10pt,stretch=1]{NL}NL & \worldflag[width=6pt,length=10pt,stretch=1]{DK}DK & 27.79 & 26.02 & 30.91 & 7.22 \\ 
   \hline
IPv6 & DFN & PowerTech & \worldflag[width=6pt,length=10pt,stretch=1]{SE}SE & \worldflag[width=6pt,length=10pt,stretch=1]{NO}NO & 27.91 & 24.83 & 31.32 & 20.32 \\ 
   \hline
IPv6 & DFN & PowerTech & \worldflag[width=6pt,length=10pt,stretch=1]{DE}DE & \worldflag[width=6pt,length=10pt,stretch=1]{SE}SE & 5.87 & 5.74 & 5.88 & 3.24 \\ 
   \hline
IPv6 & DFN & Uninett & \worldflag[width=6pt,length=10pt,stretch=1]{DE}DE & \worldflag[width=6pt,length=10pt,stretch=1]{GB}GB & 18.58 & 12.62 & 21.24 & 99.86 \\ 
   \hline
IPv6 & DFN & Uninett & \worldflag[width=6pt,length=10pt,stretch=1]{GB}GB & \worldflag[width=6pt,length=10pt,stretch=1]{DK}DK & 27.06 & 23.28 & 34.05 & 99.77 \\ 
   \hline
IPv6 & DFN & Uninett & \worldflag[width=6pt,length=10pt,stretch=1]{DK}DK & \worldflag[width=6pt,length=10pt,stretch=1]{NO}NO & 38.03 & 32.08 & 44.74 & 42.78 \\ 
   \hline
\end{tabular}
\end{adjustbox}
\end{table*}

%% file: Measurements/TEX/Traceroute-HUUDEKAU-NTNU-UDE-NTNU-AS-P2.5.tex
\begin{table*}
\centering
\caption{Inter-AS Observations Percentage ($\ge$2.5) and Mean RTT as well as 10\%-/90\%-Quantiles RTT (in ms) for Essen, DE $\rightarrow$ Trondheim, NO} 
\label{tab:AS-UDE-NTNU}
\begin{adjustbox}{max width=\textwidth}
\small
\begin{tabular}{|c|c|c|c|c|c|c|c|c|}
  \hline
{\textbf{IP}} & {\textbf{From ISP}} & {\textbf{To ISP}} & {\textbf{From}} & {\textbf{To}} & {\textbf{Mean}} & {\textbf{Q$_{\mathrm{10\%}}$}} & {\textbf{Q$_{\mathrm{90\%}}$}} & {\textbf{\%}} \\ 
  \hline
\hline
IPv4 & DFN & PowerTech &   1299: TWELVE99 &   2116: GLOBALCONNECT & 31.79 & 27.61 & 34.88 & 92.38 \\ 
   \hline
IPv4 & DFN & PowerTech &    680: DFN &   1299: TWELVE99 & 6.21 & 5.08 & 7.11 & 89.96 \\ 
   \hline
IPv4 & DFN & PowerTech &  12552: IPO-EU &   2116: GLOBALCONNECT & 28.84 & 27.21 & 30.88 & 7.24 \\ 
   \hline
IPv4 & DFN & Uninett &    680: DFN &  20965: GEANT & 6.52 & 4.97 & 7.77 & 99.82 \\ 
   \hline
IPv4 & DFN & Uninett &   2603: NORDUNET &    224: UNINETT & 32.80 & 26.41 & 47.37 & 99.78 \\ 
   \hline
IPv4 & DFN & Uninett &  20965: GEANT &   2603: NORDUNET & 23.81 & 18.39 & 39.70 & 99.75 \\ 
   \hline
IPv6 & DFN & PowerTech &   1299: TWELVE99 &   2116: GLOBALCONNECT & 27.80 & 24.96 & 32.68 & 82.47 \\ 
   \hline
IPv6 & DFN & PowerTech &    680: DFN &   1299: TWELVE99 & 6.49 & 5.59 & 7.15 & 11.85 \\ 
   \hline
IPv6 & DFN & PowerTech &  12552: IPO-EU &   2116: GLOBALCONNECT & 28.11 & 27.85 & 27.96 & 2.62 \\ 
   \hline
IPv6 & DFN & Uninett &  20965: GEANT &   2603: NORDUNET & 27.06 & 23.28 & 34.05 & 99.77 \\ 
   \hline
IPv6 & DFN & Uninett &   2603: NORDUNET &    224: UNINETT & 38.06 & 31.46 & 44.34 & 68.97 \\ 
   \hline
IPv6 & DFN & Uninett &    680: DFN &  20965: GEANT & 6.63 & 5.64 & 6.99 & 21.26 \\ 
   \hline
\end{tabular}
\end{adjustbox}
\end{table*}

%% file: Measurements/TEX/Traceroute-HUUDEKAU-NTNU-UDE-NTNU-Hops.tex
\begin{table*}
\centering
\caption{Hop count statistics for Essen, DE $\rightarrow$ Trondheim, NO} 
\label{tab:Hops-UDE-NTNU}
\begin{adjustbox}{max width=\textwidth}
\small
\begin{tabular}{|c|c|c|c|c|c|c|c|}
  \hline
{\textbf{IP}} & {\textbf{From ISP}} & {\textbf{To ISP}} & {\textbf{Min}} & {\textbf{Q$_{\mathrm{10\%}}$}} & {\textbf{Mean}} & {\textbf{Median}} & {\textbf{Q$_{\mathrm{90\%}}$}} \\ 
  \hline
\hline
IPv4 & DFN & PowerTech &  11 & 13.00 & 13.31 & 13.00 & 14.00 \\ 
   \hline
IPv4 & DFN & Uninett &  16 & 17.00 & 17.81 & 17.00 & 19.00 \\ 
   \hline
IPv6 & DFN & PowerTech &  11 & 11.00 & 12.63 & 13.00 & 14.00 \\ 
   \hline
IPv6 & DFN & Uninett &  16 & 17.00 & 17.87 & 17.00 & 19.00 \\ 
   \hline
\end{tabular}
\end{adjustbox}
\end{table*}

%% file: Measurements/TEX/Traceroute-HUUDEKAU-NTNU-HU-NTNU-Country-P15.tex
\begin{table*}
\centering
\caption{Inter-Country Observations Percentage ($\ge$15.0) and Mean RTT as well as 10\%-/90\%-Quantiles RTT (in ms) for Haikou, CN $\rightarrow$ Trondheim, NO} 
\label{tab:Country-HU-NTNU}
\begin{adjustbox}{max width=\textwidth}
\small
\begin{tabular}{|c|c|c|c|c|c|c|c|c|}
  \hline
{\textbf{IP}} & {\textbf{From ISP}} & {\textbf{To ISP}} & {\textbf{From}} & {\textbf{To}} & {\textbf{Mean}} & {\textbf{Q$_{\mathrm{10\%}}$}} & {\textbf{Q$_{\mathrm{90\%}}$}} & {\textbf{\%}} \\ 
  \hline
\hline
IPv4 & CERNET & PowerTech & \worldflag[width=6pt,length=10pt,stretch=1]{US}US & \worldflag[width=6pt,length=10pt,stretch=1]{NO}NO & 376.20 & 335.06 & 438.30 & 75.34 \\ 
   \hline
IPv4 & CERNET & PowerTech & \worldflag[width=6pt,length=10pt,stretch=1]{CN}CN & \worldflag[width=6pt,length=10pt,stretch=1]{US}US & 207.92 & 190.39 & 243.54 & 61.68 \\ 
   \hline
IPv4 & CERNET & PowerTech & \worldflag[width=6pt,length=10pt,stretch=1]{CN}CN & \worldflag[width=6pt,length=10pt,stretch=1]{CN}HK & 87.07 & 79.22 & 98.21 & 36.51 \\ 
   \hline
IPv4 & CERNET & PowerTech & \worldflag[width=6pt,length=10pt,stretch=1]{CN}HK & \worldflag[width=6pt,length=10pt,stretch=1]{US}US & 222.75 & 202.96 & 254.01 & 32.83 \\ 
   \hline
IPv4 & CERNET & PowerTech & \worldflag[width=6pt,length=10pt,stretch=1]{US}US & \worldflag[width=6pt,length=10pt,stretch=1]{CN}HK & 225.88 & 215.85 & 244.90 & 31.78 \\ 
   \hline
IPv4 & CERNET & Uninett & \worldflag[width=6pt,length=10pt,stretch=1]{GB}GB & \worldflag[width=6pt,length=10pt,stretch=1]{SE}SE & 314.55 & 291.48 & 336.77 & 80.12 \\ 
   \hline
IPv4 & CERNET & Uninett & \worldflag[width=6pt,length=10pt,stretch=1]{SE}SE & \worldflag[width=6pt,length=10pt,stretch=1]{NO}NO & 333.69 & 306.62 & 371.39 & 75.93 \\ 
   \hline
IPv4 & CERNET & Uninett & \worldflag[width=6pt,length=10pt,stretch=1]{CN}CN & \worldflag[width=6pt,length=10pt,stretch=1]{GB}GB & 276.58 & 269.81 & 302.09 & 74.46 \\ 
   \hline
IPv4 & CERNET & Uninett & \worldflag[width=6pt,length=10pt,stretch=1]{DK}DK & \worldflag[width=6pt,length=10pt,stretch=1]{NO}NO & 316.31 & 295.05 & 351.97 & 23.00 \\ 
   \hline
IPv4 & CERNET & Uninett & \worldflag[width=6pt,length=10pt,stretch=1]{SE}SE & \worldflag[width=6pt,length=10pt,stretch=1]{DK}DK & 307.45 & 285.60 & 330.72 & 17.04 \\ 
   \hline
IPv4 & CnUnicom & PowerTech & \worldflag[width=6pt,length=10pt,stretch=1]{SE}SE & \worldflag[width=6pt,length=10pt,stretch=1]{NO}NO & 246.92 & 197.29 & 349.83 & 89.13 \\ 
   \hline
IPv4 & CnUnicom & PowerTech & \worldflag[width=6pt,length=10pt,stretch=1]{DE}DE & \worldflag[width=6pt,length=10pt,stretch=1]{SE}SE & 214.23 & 174.91 & 253.12 & 51.08 \\ 
   \hline
IPv4 & CnUnicom & PowerTech & \worldflag[width=6pt,length=10pt,stretch=1]{CN}CN & \worldflag[width=6pt,length=10pt,stretch=1]{GB}GB & 215.33 & 193.85 & 253.07 & 33.60 \\ 
   \hline
IPv4 & CnUnicom & PowerTech & \worldflag[width=6pt,length=10pt,stretch=1]{CN}CN & \worldflag[width=6pt,length=10pt,stretch=1]{US}US & 218.88 & 184.65 & 266.47 & 32.44 \\ 
   \hline
IPv4 & CnUnicom & PowerTech & \worldflag[width=6pt,length=10pt,stretch=1]{GB}GB & \worldflag[width=6pt,length=10pt,stretch=1]{SE}SE & 228.30 & 210.22 & 246.21 & 26.43 \\ 
   \hline
IPv4 & CnUnicom & PowerTech & \worldflag[width=6pt,length=10pt,stretch=1]{SE}SE & \worldflag[width=6pt,length=10pt,stretch=1]{DE}DE & 232.66 & 210.07 & 265.85 & 26.33 \\ 
   \hline
IPv4 & CnUnicom & PowerTech & \worldflag[width=6pt,length=10pt,stretch=1]{CN}CN & \worldflag[width=6pt,length=10pt,stretch=1]{DE}DE & 206.06 & 186.35 & 232.84 & 15.64 \\ 
   \hline
IPv4 & CnUnicom & Uninett & \worldflag[width=6pt,length=10pt,stretch=1]{SE}SE & \worldflag[width=6pt,length=10pt,stretch=1]{NO}NO & 307.06 & 275.14 & 362.59 & 71.87 \\ 
   \hline
IPv4 & CnUnicom & Uninett & \worldflag[width=6pt,length=10pt,stretch=1]{CN}CN & \worldflag[width=6pt,length=10pt,stretch=1]{US}US & 247.41 & 197.88 & 293.88 & 53.82 \\ 
   \hline
IPv4 & CnUnicom & Uninett & \worldflag[width=6pt,length=10pt,stretch=1]{GB}GB & \worldflag[width=6pt,length=10pt,stretch=1]{SE}SE & 306.65 & 235.80 & 380.48 & 47.47 \\ 
   \hline
IPv4 & CnUnicom & Uninett & \worldflag[width=6pt,length=10pt,stretch=1]{NL}NL & \worldflag[width=6pt,length=10pt,stretch=1]{DK}DK & 247.77 & 179.97 & 302.28 & 43.63 \\ 
   \hline
IPv4 & CnUnicom & Uninett & \worldflag[width=6pt,length=10pt,stretch=1]{US}US & \worldflag[width=6pt,length=10pt,stretch=1]{GB}GB & 302.04 & 256.71 & 351.67 & 38.66 \\ 
   \hline
IPv4 & CnUnicom & Uninett & \worldflag[width=6pt,length=10pt,stretch=1]{DK}DK & \worldflag[width=6pt,length=10pt,stretch=1]{SE}SE & 289.73 & 231.55 & 340.19 & 32.66 \\ 
   \hline
IPv4 & CnUnicom & Uninett & \worldflag[width=6pt,length=10pt,stretch=1]{DE}DE & \worldflag[width=6pt,length=10pt,stretch=1]{NL}NL & 191.01 & 162.81 & 223.90 & 24.80 \\ 
   \hline
IPv4 & CnUnicom & Uninett & \worldflag[width=6pt,length=10pt,stretch=1]{US}US & \worldflag[width=6pt,length=10pt,stretch=1]{NL}NL & 291.89 & 264.12 & 343.71 & 23.74 \\ 
   \hline
IPv4 & CnUnicom & Uninett & \worldflag[width=6pt,length=10pt,stretch=1]{DK}DK & \worldflag[width=6pt,length=10pt,stretch=1]{NO}NO & 210.20 & 178.01 & 257.98 & 22.35 \\ 
   \hline
IPv4 & CnUnicom & Uninett & \worldflag[width=6pt,length=10pt,stretch=1]{CN}CN & \worldflag[width=6pt,length=10pt,stretch=1]{DE}DE & 190.05 & 171.85 & 223.94 & 15.88 \\ 
   \hline
IPv6 & CERNET & PowerTech & \worldflag[width=6pt,length=10pt,stretch=1]{SE}SE & \worldflag[width=6pt,length=10pt,stretch=1]{NO}NO & 347.19 & 284.01 & 391.75 & 94.67 \\ 
   \hline
IPv6 & CERNET & PowerTech & \worldflag[width=6pt,length=10pt,stretch=1]{US}US & \worldflag[width=6pt,length=10pt,stretch=1]{SE}SE & 338.20 & 277.28 & 383.12 & 93.62 \\ 
   \hline
IPv6 & CERNET & PowerTech & \worldflag[width=6pt,length=10pt,stretch=1]{CN}HK & \worldflag[width=6pt,length=10pt,stretch=1]{US}US & 136.07 & 112.89 & 199.07 & 28.82 \\ 
   \hline
IPv6 & CERNET & PowerTech & \worldflag[width=6pt,length=10pt,stretch=1]{US}US & \worldflag[width=6pt,length=10pt,stretch=1]{GB}GB & 333.96 & 324.42 & 348.03 & 26.82 \\ 
   \hline
IPv6 & CERNET & PowerTech & \worldflag[width=6pt,length=10pt,stretch=1]{US}US & \worldflag[width=6pt,length=10pt,stretch=1]{IE}IE & 323.74 & 213.60 & 395.62 & 21.75 \\ 
   \hline
IPv6 & CERNET & PowerTech & \worldflag[width=6pt,length=10pt,stretch=1]{GB}GB & \worldflag[width=6pt,length=10pt,stretch=1]{NL}NL & 337.45 & 330.81 & 343.40 & 18.51 \\ 
   \hline
IPv6 & CERNET & Uninett & \worldflag[width=6pt,length=10pt,stretch=1]{DK}DK & \worldflag[width=6pt,length=10pt,stretch=1]{NO}NO & 346.60 & 316.79 & 389.20 & 42.47 \\ 
   \hline
IPv6 & CERNET & Uninett & \worldflag[width=6pt,length=10pt,stretch=1]{SE}SE & \worldflag[width=6pt,length=10pt,stretch=1]{DK}DK & 362.50 & 315.02 & 435.22 & 38.07 \\ 
   \hline
IPv6 & CERNET & Uninett & \worldflag[width=6pt,length=10pt,stretch=1]{CN}HK & \worldflag[width=6pt,length=10pt,stretch=1]{US}US & 163.30 & 80.02 & 258.21 & 29.69 \\ 
   \hline
IPv6 & CERNET & Uninett & \worldflag[width=6pt,length=10pt,stretch=1]{GB}GB & \worldflag[width=6pt,length=10pt,stretch=1]{DK}DK & 303.43 & 275.64 & 326.74 & 24.74 \\ 
   \hline
IPv6 & CERNET & Uninett & \worldflag[width=6pt,length=10pt,stretch=1]{US}US & \worldflag[width=6pt,length=10pt,stretch=1]{SE}SE & 197.52 & 129.53 & 275.31 & 23.40 \\ 
   \hline
IPv6 & CERNET & Uninett & \worldflag[width=6pt,length=10pt,stretch=1]{US}US & \worldflag[width=6pt,length=10pt,stretch=1]{DK}DK & 323.45 & 257.75 & 355.29 & 21.80 \\ 
   \hline
IPv6 & CERNET & Uninett & \worldflag[width=6pt,length=10pt,stretch=1]{CN}HK & \worldflag[width=6pt,length=10pt,stretch=1]{GB}GB & 278.39 & 273.17 & 280.65 & 18.18 \\ 
   \hline
IPv6 & CnUnicom & PowerTech & \worldflag[width=6pt,length=10pt,stretch=1]{SE}SE & \worldflag[width=6pt,length=10pt,stretch=1]{NO}NO & 307.82 & 289.45 & 323.36 & 29.15 \\ 
   \hline
IPv6 & CnUnicom & PowerTech & \worldflag[width=6pt,length=10pt,stretch=1]{CN}CN & \worldflag[width=6pt,length=10pt,stretch=1]{US}US & 258.21 & 197.77 & 297.99 & 27.67 \\ 
   \hline
IPv6 & CnUnicom & Uninett & \worldflag[width=6pt,length=10pt,stretch=1]{DK}DK & \worldflag[width=6pt,length=10pt,stretch=1]{NO}NO & 309.66 & 293.72 & 336.73 & 97.65 \\ 
   \hline
IPv6 & CnUnicom & Uninett & \worldflag[width=6pt,length=10pt,stretch=1]{CN}CN & \worldflag[width=6pt,length=10pt,stretch=1]{US}US & 249.08 & 196.61 & 298.16 & 35.87 \\ 
   \hline
\end{tabular}
\end{adjustbox}
\end{table*}

%% file: Measurements/TEX/Traceroute-HUUDEKAU-NTNU-HU-NTNU-AS-P10.tex
\begin{table*}
\centering
\caption{Inter-AS Observations Percentage ($\ge$10.0) and Mean RTT as well as 10\%-/90\%-Quantiles RTT (in ms) for Haikou, CN $\rightarrow$ Trondheim, NO} 
\label{tab:AS-HU-NTNU}
\begin{adjustbox}{max width=\textwidth}
\small
\begin{tabular}{|c|c|c|c|c|c|c|c|c|}
  \hline
{\textbf{IP}} & {\textbf{From ISP}} & {\textbf{To ISP}} & {\textbf{From}} & {\textbf{To}} & {\textbf{Mean}} & {\textbf{Q$_{\mathrm{10\%}}$}} & {\textbf{Q$_{\mathrm{90\%}}$}} & {\textbf{\%}} \\ 
  \hline
\hline
IPv4 & CERNET & PowerTech &   3356: LEVEL3 &   2116: GLOBALCONNECT & 375.19 & 331.92 & 449.70 & 84.59 \\ 
   \hline
IPv4 & CERNET & PowerTech &   4538: ERX-CERNET-BKB &    174: COGENT-174 & 207.56 & 190.42 & 242.74 & 59.19 \\ 
   \hline
IPv4 & CERNET & PowerTech &   4538: ERX-CERNET-BKB &   4637: ASN-TELSTRA & 86.90 & 79.39 & 98.05 & 34.40 \\ 
   \hline
IPv4 & CERNET & PowerTech &   4637: ASN-TELSTRA &   3356: LEVEL3 & 387.59 & 347.33 & 431.54 & 33.54 \\ 
   \hline
IPv4 & CERNET & PowerTech &    174: COGENT-174 &   2116: GLOBALCONNECT & 374.38 & 362.29 & 398.42 & 10.13 \\ 
   \hline
IPv4 & CERNET & Uninett &   2603: NORDUNET &    224: UNINETT & 329.68 & 299.18 & 368.26 & 98.93 \\ 
   \hline
IPv4 & CERNET & Uninett &   4538: ERX-CERNET-BKB &  23911: CNGI-BJIX-AS & 50.96 & 45.29 & 67.58 & 90.04 \\ 
   \hline
IPv4 & CERNET & Uninett &  20965: GEANT &   2603: NORDUNET & 309.99 & 280.11 & 329.92 & 82.62 \\ 
   \hline
IPv4 & CERNET & Uninett &  23911: CNGI-BJIX-AS &  20965: GEANT & 276.60 & 270.18 & 302.09 & 74.42 \\ 
   \hline
IPv4 & CERNET & Uninett &  23911: CNGI-BJIX-AS &  11537: INTERNET2 & 251.56 & 212.39 & 282.16 & 13.24 \\ 
   \hline
IPv4 & CERNET & Uninett &  11537: INTERNET2 &   2603: NORDUNET & 298.91 & 263.60 & 347.77 & 12.98 \\ 
   \hline
IPv4 & CnUnicom & PowerTech &   1299: TWELVE99 &   2116: GLOBALCONNECT & 256.48 & 198.43 & 363.89 & 94.84 \\ 
   \hline
IPv4 & CnUnicom & PowerTech &   4837: CHINA169-BB &   1299: TWELVE99 & 215.46 & 186.53 & 265.50 & 87.81 \\ 
   \hline
IPv4 & CnUnicom & Uninett &   2603: NORDUNET &    224: UNINETT & 284.06 & 193.62 & 357.59 & 94.22 \\ 
   \hline
IPv4 & CnUnicom & Uninett &   1299: TWELVE99 &   2603: NORDUNET & 264.07 & 183.05 & 337.43 & 87.89 \\ 
   \hline
IPv4 & CnUnicom & Uninett &   4837: CHINA169-BB &   1299: TWELVE99 & 231.80 & 176.31 & 282.20 & 76.16 \\ 
   \hline
IPv6 & CERNET & PowerTech & 138440: CNGI-HAK-IX-AS &  23910: CNGI-CERNET2-AS & 2.43 & 1.49 & 3.52 & 70.26 \\ 
   \hline
IPv6 & CERNET & PowerTech & 138440: CNGI-HAK-IX-AS &  24357: CNGI-GZ-IX-AS & 16.87 & 15.01 & 17.66 & 26.66 \\ 
   \hline
IPv6 & CERNET & PowerTech &  24357: CNGI-GZ-IX-AS &  23910: CNGI-CERNET2-AS & 11.63 & 10.64 & 12.08 & 26.62 \\ 
   \hline
IPv6 & CERNET & PowerTech &  23910: CNGI-CERNET2-AS &  23911: CNGI-BJIX-AS & 48.99 & 44.62 & 49.14 & 14.18 \\ 
   \hline
IPv6 & CERNET & Uninett & 138440: CNGI-HAK-IX-AS &  23910: CNGI-CERNET2-AS & 2.41 & 1.47 & 3.54 & 70.26 \\ 
   \hline
IPv6 & CERNET & Uninett &   2603: NORDUNET &    224: UNINETT & 355.21 & 301.21 & 394.09 & 67.58 \\ 
   \hline
IPv6 & CERNET & Uninett &   1299: TWELVE99 &   2603: NORDUNET & 351.75 & 276.71 & 434.60 & 44.11 \\ 
   \hline
IPv6 & CERNET & Uninett & 138440: CNGI-HAK-IX-AS &  24357: CNGI-GZ-IX-AS & 16.83 & 14.93 & 17.68 & 26.66 \\ 
   \hline
IPv6 & CERNET & Uninett &  24357: CNGI-GZ-IX-AS &  23910: CNGI-CERNET2-AS & 11.64 & 10.64 & 12.09 & 26.62 \\ 
   \hline
IPv6 & CERNET & Uninett &   6939: HURRICANE &   1299: TWELVE99 & 197.42 & 129.53 & 274.54 & 23.40 \\ 
   \hline
IPv6 & CERNET & Uninett &  20965: GEANT &   2603: NORDUNET & 309.72 & 292.95 & 327.06 & 18.70 \\ 
   \hline
IPv6 & CERNET & Uninett &  23911: CNGI-BJIX-AS &  20965: GEANT & 278.39 & 273.17 & 280.65 & 18.18 \\ 
   \hline
IPv6 & CERNET & Uninett &  11537: INTERNET2 &   2603: NORDUNET & 333.14 & 312.55 & 354.72 & 16.47 \\ 
   \hline
IPv6 & CERNET & Uninett &  23910: CNGI-CERNET2-AS &  23911: CNGI-BJIX-AS & 49.09 & 44.62 & 49.20 & 14.22 \\ 
   \hline
IPv6 & CERNET & Uninett &  23911: CNGI-BJIX-AS &  11537: INTERNET2 & 248.50 & 210.79 & 258.93 & 11.20 \\ 
   \hline
IPv6 & CnUnicom & PowerTech &   1299: TWELVE99 &   2116: GLOBALCONNECT & 307.94 & 289.50 & 325.78 & 32.29 \\ 
   \hline
IPv6 & CnUnicom & PowerTech &   4837: CHINA169-BB &    701: UUNET & 251.90 & 197.04 & 298.06 & 22.11 \\ 
   \hline
IPv6 & CnUnicom & PowerTech &   3356: LEVEL3 &   2116: GLOBALCONNECT & 301.29 & 288.63 & 329.44 & 13.29 \\ 
   \hline
IPv6 & CnUnicom & PowerTech &    701: UUNET &   1299: TWELVE99 & 265.27 & 240.60 & 314.46 & 11.34 \\ 
   \hline
IPv6 & CnUnicom & Uninett &   2603: NORDUNET &    224: UNINETT & 309.66 & 293.72 & 336.73 & 97.65 \\ 
   \hline
IPv6 & CnUnicom & Uninett &   4837: CHINA169-BB &    701: UUNET & 249.08 & 196.61 & 298.16 & 35.87 \\ 
   \hline
IPv6 & CnUnicom & Uninett &    701: UUNET &   1299: TWELVE99 & 256.06 & 237.25 & 277.77 & 14.69 \\ 
   \hline
\end{tabular}
\end{adjustbox}
\end{table*}

%% file: Measurements/TEX/Traceroute-HUUDEKAU-NTNU-HU-NTNU-Hops.tex
\begin{table*}
\centering
\caption{Hop count statistics for Haikou, CN $\rightarrow$ Trondheim, NO} 
\label{tab:Hops-HU-NTNU}
\begin{adjustbox}{max width=\textwidth}
\small
\begin{tabular}{|c|c|c|c|c|c|c|c|}
  \hline
{\textbf{IP}} & {\textbf{From ISP}} & {\textbf{To ISP}} & {\textbf{Min}} & {\textbf{Q$_{\mathrm{10\%}}$}} & {\textbf{Mean}} & {\textbf{Median}} & {\textbf{Q$_{\mathrm{90\%}}$}} \\ 
  \hline
\hline
IPv4 & CERNET & PowerTech &  15 & 21.00 & 24.18 & 24.00 & 31.00 \\ 
   \hline
IPv4 & CERNET & Uninett &  19 & 24.00 & 26.51 & 25.00 & 31.00 \\ 
   \hline
IPv4 & CnUnicom & PowerTech &  14 & 16.00 & 16.84 & 17.00 & 18.00 \\ 
   \hline
IPv4 & CnUnicom & Uninett &  13 & 21.00 & 23.13 & 24.00 & 25.00 \\ 
   \hline
IPv6 & CERNET & PowerTech &  20 & 22.00 & 26.91 & 28.00 & 30.00 \\ 
   \hline
IPv6 & CERNET & Uninett &  19 & 24.00 & 28.96 & 30.00 & 31.00 \\ 
   \hline
\end{tabular}
\end{adjustbox}
\end{table*}